\documentclass[aps,prd,twocolumn,nofootinbib,superscriptaddress]{revtex4-2}

\usepackage{slashed}
\usepackage{amssymb}
\usepackage{bm}
\usepackage{graphicx}
\usepackage{hyperref}
\usepackage{mathrsfs}

\usepackage{amsmath}

\newcommand{\sumint}{%
  \mathop{\ooalign{%
    $\displaystyle\sum$\cr
    \hidewidth$\displaystyle\int$\hidewidth\cr
  }}\nolimits
}

\usepackage{xcolor}

\newcommand{\beq}{\begin{equation}}
\newcommand{\eeq}{\end{equation}}
\newcommand{\bqa}{\begin{eqnarray}}
\newcommand{\eqa}{\end{eqnarray}}

\newcommand{\Nc}{N_c}
\newcommand{\dA}{d_A}
\newcommand{\Tr}{\mathrm{Tr}}

\begin{document}

\title{Two-loop HTL-resummed thermodynamics for pure
  $\mathcal N=2$ supersymmetric Yang--Mills theory}

\author{Qianqian Du}
\email{dqianqian@mailbox.gxnu.edu.cn}
\affiliation{Guangxi Key Laboratory of Nuclear Physics and Technology, Guangxi Normal University, Guilin 541004, China}
\affiliation{Department of Physics, Guangxi Normal University, Guilin, 541004, China}

\author{Bingxiang Luo}
\affiliation{Guangxi Key Laboratory of Nuclear Physics and Technology, Guangxi Normal University, Guilin 541004, China}
\affiliation{Department of Physics, Guangxi Normal University, Guilin, 541004, China}

\author{Ubaid Tantary}
\email{utantary@pmu.edu.sa}
\affiliation{Department of Mathematics and Natural Sciences,
Prince Mohammad Bin Fahd University, Al Khobar 31952, Saudi Arabia}
\date{\today}

\begin{abstract}
We compute the two-loop hard-thermal-loop (HTL) resummed thermodynamic potential of pure four-dimensional $\mathcal N=2$ supersymmetric Yang-Mills theory using regularization by dimensional reduction. Our result is gauge invariant and was renormalized using vacuum energy, gluon mass $m_d$ and quark mass $m_q$ counterterms. The thermal mass parameters are determined self-consistently and numerically from gap equations incorporating the running coupling. Based on these solutions, we numerically calculate the scaled pressure, entropy density, energy density, and trace anomaly and compare them with the results in the static perturbative limit through $\mathcal O(\lambda^2)$. For the first three quantities, next-to-leading-order (NLO) corrections are positive and have discernible effects at small coupling, and reduce the renormalization-scale dependence at relatively large couplings.
We assess $m_q$-prescription dependence by comparing these results with those obtained by setting $m_q=0$. 
We find that it has a substantial impact on $m_d$ and the scalar mass $m_s$ at large couplings, and it reverses the signs of the NLO corrections to the first three quantities and qualitatively alters the trace anomaly. 
 
\end{abstract}

\maketitle

\section{Introduction}
\label{sec:introduction}

Hard-thermal-loop perturbation theory (HTLpt) reorganizes thermal
perturbation theory by adding and subtracting the gauge-invariant HTL effective action~\cite{Braaten:1991gm,Andersen:1999fw,Andersen:1999sf}.
This reorganization incorporates HTL self-energies into the propagators and includes the corresponding HTL vertices, while treating the subtraction terms perturbatively to avoid double counting. It incorporates thermal screening effects while preserving gauge invariance, without introducing a local gluon mass term that would violate gauge invariance. This framework was applied to QCD thermodynamics at two-loop order~\cite{Andersen:2002ey,Andersen:2003zk} and three-loop order~\cite{Andersen:2010ct,Andersen:2011sf,Andersen:2010wu} .

The two-loop HTLpt thermodynamic potential has also been computed for $\mathcal N=4$ supersymmetric Yang--Mills theory (${\rm SYM}_{4,4}$)
using the canonical dimensional regularization (DRG) scheme \cite{Du:2020odw} and  the regularization by dimensional reduction (RDR) scheme which preserves
supersymmetry~\cite{Du:2022uyw}, which indicates that different regularization schemes lead to only minor numerical changes, while the coefficients of $\lambda$, $\lambda^{3/2}$, and $\lambda^2\log\lambda$ terms are scheme independent.
In this paper, we focus on the HTLpt calculation of the pure $\mathcal N=2$ supersymmetric Yang--Mills theory (${\rm SYM}_{2,4}$) in the RDR scheme. 
Although the strict weak-coupling free energy of pure ${\rm SYM}_{2,4}$ has been computed up to order $\lambda^2$~\cite{Tantary:2026udn}, the two-loop HTLpt of ${\rm SYM}_{2,4}$ reorganizes the perturbative expansion through thermal-mass resummation, providing a complementary approach to studying its
thermodynamics at finite coupling.
The thermal mass parameters $m_d$, $m_s$ and $m_q$ are determined self-consistently by using the variational method, and the unexpanded expressions retain selected higher-order contributions.
Although the two-loop HTLpt does not generate complete $\mathcal O(\lambda^2)$ terms, it allows us to investigate the effects of thermal masses, expansion order, renormalization scale, and quark mass prescription on thermodynamic functions.
  
For pure ${\rm SYM}_{2,4}$, there are two adjoint Majorana fermions and two real adjoint scalars rather than the four and six of ${\rm SYM}_{4,4}$, respectively. This leads to different diagrammatic contributions to the thermodynamic potential and the coupled gap equations.
In addition, the adjoint 't Hooft coupling $\lambda=g^2N_c$ runs, and satisfies
\beq
\mu\frac{d\lambda}{d\mu}
 =-\frac{\lambda^2}{4\pi^2}+O(\lambda^3),
\label{eq:beta}
\eeq
where $g$ is the gauge coupling, $N_c$ is the number of colors, and $\mu$ is the renormalization scale. As a result, the nonzero beta function introduces additional scale dependence. 
Consequently, the normalized pressure, entropy density, and energy density are not the same, and a nonvanishing trace anomaly appears.
In addition, in SYM$_{4,4}$, although setting $m_q=0$ leads to quantitative rather than qualitative changes for $m_d$, $m_s$ and thermodynamic functions, whether this remains true for SYM$_{2,4}$ requires further investigation.
These motivate us to calculate the two-loop HTLpt result of SYM$_{2,4}$ to investigate the self-consistent thermal masses, and the dependence of the thermodynamic functions on the expansion order, renormalization scale, and quark mass prescription.

The paper is organized as follows. In Section~\ref{sec:theory}, we briefly introduce the basics of the HTLpt framework of SYM$_{2,4}$, and give the gluon, fermion, and scalar self-energies in the HTL approximation, discuss their relation to the corresponding results in SYM$_{4,4}$, and determine their leading-order thermal masses. 
In Section~\ref{sec:potential}, we explore the relation of the one- and two-loop free energy in HTLpt between SYM$_{2,4}$ and SYM$_{4,4}$, calculate the leading-order (LO) and NLO HTLpt thermodynamic potential, and determine the vacuum-energy, gluon mass, and fermion mass counterterms.
The numerical solutions of the gap equation  incorporating the solution of Eq.~(\ref{eq:beta})
are shown in Sec.~\ref{sec:numerics}, we also discuss the effects of setting $m_q=0$ on $m_d$ and $m_s$ in the meantime.
Based on Sec.~\ref{sec:numerics}, we calculate the scaled pressure, energy density, entropy density, and trace anomaly numerically in Sec.~\ref{numerics-themodynamics}, and compare with the strict results up to order $\lambda^2$ and the corresponding results with $m_q=0$.
Sec.~\ref{sec:summary} is the summary and outlook. 

\section{ The basics of the HTLpt for pure $\mathcal N=2$ SYM}
\label{sec:theory}

The Lagrangian in Minkowski space with real-scalar field and Majorana-fermions is~\cite{Tantary:2026udn},
\begin{align}
\mathcal L_{\rm SYM}
={}&\Tr\biggl[
 -\frac12 G_{\mu\nu}G^{\mu\nu}
 +(D_\mu\Phi_I)(D^\mu\Phi_I)
 +i\bar\psi_i\!\not\!D\,\psi_i
\nonumber\\
&
 -\frac12 g^2\bigl(i[\Phi_I,\Phi_J]\bigr)^2
 -ig\bar\psi_i
 \bigl[\alpha_{ij}X+i\beta_{ij}\gamma_5Y,\psi_j\bigr]
 \biggr]
\nonumber\\
&+\mathcal L_{\rm gf}+\mathcal L_{\rm gh}
 +\Delta\mathcal L_{\rm SYM},
\label{eq:lagrangian}
\end{align}
where $\Phi=(X,Y)$, $X$ and $Y$ are Hermitian and denote scalar and pseudoscalar fields. $I=1,2$ label the components of the vector $\Phi$. The Latin indices $i=1,2$ are used to label the two Majorana fermions, with $\psi_i$ denoting each bispinor.  All fields are in the adjoint representation of $SU(\Nc)$.  The $2\times2$ matrices $\alpha$ and $\beta$ satisfy 
\beq
\label{alpha-beta-matrix}
\alpha_{ij}\alpha_{jk} = -\delta_{ik},\quad\quad
\beta_{ij} \beta_{ji} = -\delta_{ii}= -2. 
\eeq
Comparing with ${\rm SYM}_{4,4}$, there is an extra factor of ``3" on the right-hand side of the first equation since there are three scalar-pseudoscalar pairs, and $\delta_{ii}=4$ for four Majorana fermions. Eq.~(\ref{alpha-beta-matrix}) accounts for the main difference between the results for these two theories.  

The four-dimensional field content follows from SUSY dimensional
reduction of six-dimensional $\mathcal N=1$ SYM.  For ultraviolet
regularization, all momentum-space integrals in RDR are evaluated in
$d=4-2\epsilon$ dimensions, while the gauge and fermion field
representation sizes and the field multiplicities are kept at their
integer $D=4$ values
\cite{Du:2022uyw}.  This is the same prescription used for the strict
weak-coupling result in Ref.~\cite{Tantary:2026udn}.  Ultraviolet poles are
removed by modified minimal subtraction in RDR; the resulting
$\overline{\rm DR}$ coupling is denoted by $\lambda(\mu)$ below.

The HTLpt Lagrangian is
\beq
 {\cal L}
 =\left.
 \left({\cal L}_{\rm SYM}+{\cal L}_{\rm HTL}\right)
 \right|_{g\rightarrow\sqrt{\delta}\,g}
 +\Delta{\cal L}_{\rm HTL},
\label{eq:delta-lagrangian}
\eeq
where
\begin{align}
{\cal L}_{\rm HTL}
={}&-\frac12(1-\delta)m_d^2
\Tr\left[
G_{\mu\alpha}
\left\langle
\frac{y^\alpha y^\beta}{(y\cdot D)^2}
\right\rangle_{\hat{\bm y}}
G_\beta^{\ \mu}
\right]
\nonumber\\
&+(1-\delta)i m_q^2
\Tr\left[
\bar\psi_i\gamma^\mu
\left\langle\frac{y_\mu}{y\cdot D}\right\rangle_{\hat{\bm y}}
\psi_i
\right]
\nonumber\\
&-(1-\delta)m_s^2\Tr(\Phi_I^2).
\label{eq:htl-lagrangian}
\end{align}
Here $y^\mu=(1,\hat{\bm y})$ is lightlike and the brackets denote an
average over the direction of $\hat{\bm y}$.  The mass parameters $m_d$, $m_s$, and $m_q$ are the gauge boson, scalar, and fermion HTL thermal masses, respectively.  HTLpt is defined by expanding in the formal parameter
$\delta$, truncating the expansion, and then setting $\delta=1$.

\begin{table}[t]
\centering
\caption{Relative weights of the one-loop bosonic self-energy in
SYM$_{2,4}$ compared with SYM$_{4,4}$.. Curly line, solid line, dashed line and dotted line represent a gluon field, fermion field, scalar field and ghost field, respectively.}
\label{gluon-self-energy-weights}
\begingroup
\setlength{\tabcolsep}{10pt}
\renewcommand{\arraystretch}{1.5}
\def\diagcell#1{%
  \raisebox{-0.5\height}{%
    \includegraphics[width=0.1\textwidth]{diagram_#1.pdf}%
  }%
}
\resizebox{\columnwidth}{!}{%
\begin{tabular}{c|c}
\hline
Self-energy diagram
&
$\Pi_{\mu\nu}^{\mathcal{N}=2}/
 \Pi_{\mu\nu}^{\mathcal{N}=4}$
\\[6pt]
\hline
\diagcell{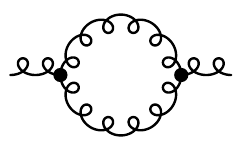}\quad
\diagcell{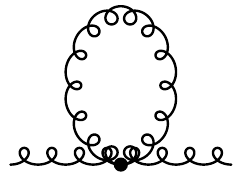}\quad
\diagcell{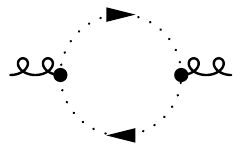}
& $1$
\\[10pt]
\diagcell{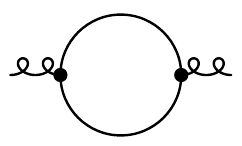}
& $\displaystyle\frac{1}{2}$
\\[10pt]
\diagcell{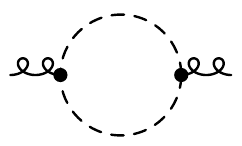}\qquad
\diagcell{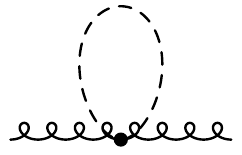}
& $\displaystyle\frac{1}{3}$
\\[10pt]
\hline
\end{tabular}
}
\endgroup
\end{table}
In HTL perturbation theory, although the three mass parameters are determined by using the variational method, their leading-order perturbative values are still needed as benchmarks for the weak-coupling analysis and numerical comparisons.
The leading-order perturbative masses can be obtained from the one-loop self-energy in the HTL limit. The gluon polarization tensor is 
\begin{align}
\Pi_{ab}^{\mu\nu}(p)&=
-\lambda\delta_{ab}\int\frac{d^3q}{(2\pi)^3}
\frac{f(\mathbf{q})}{|\mathbf{q}|}\frac{1}{(p\cdot q)^2} \nonumber\\&\times\big[
p^2q^\mu q^\nu-(p\cdot q)(p^\mu q^\nu+q^\mu p^\nu)
+(p\cdot q)^2g^{\mu\nu}\big],
\label{HTL-energy-gluon}
\end{align}
where $f(\mathbf{q})=2n_g(\mathbf{q})+4n_f(\mathbf{q})+2n_s(\mathbf{q})$ is the effective one-particle distribution function for ${\cal N}=2$ SYM. $n_{g,s}(\mathbf{q})=1/(e^{{\mathbf{|q|}}/T}-1)$ is the bosonic distribution function, and $n_{f}(\mathbf{q})=1/(e^{{\mathbf{|q|}}/T}+1)$ is the fermionic distribution function. The coefficients of $n_g$, $n_f$, $n_s$ correspond to the number of degrees of freedom of the gauge field, fermions, and scalars. The Debye mass for the gauge field can be obtained by 
\beq
m_{d,{\rm LO}}^2
=-g_{\mu\nu}\Pi^{\mu\nu}(p)
=2\lambda\int\frac{d^3q}{(2\pi)^3}
\frac{f(\mathbf{q})}{|\mathbf{q}|}
=\lambda T^2.
\label{mass-md}
\eeq
The difference of $\Pi_{ab}^{\mu\nu}(p)$ between SYM$_{2,4}$ and SYM$_{4,4}$ in Ref.~\cite{Czajka:2012gq} is only reflected in $f(\mathbf{q})$; for the latter case, the coefficients of $n_g$, $n_f$, $n_s$ are ``2'', ``8'' and ``6'', which are the number of degrees of freedom for ${\cal N}=4$. These changes are shown in Tab.~\ref{gluon-self-energy-weights}, by using Eq.~(\ref{alpha-beta-matrix}).
The HTL scalar self-energy is
\beq
\mathcal{P}_{s,ab}^{IJ}=
g^2N_c\delta_{ab}\delta^{IJ}
\int\frac{d^3q}{(2\pi)^3}
\frac{f(\mathbf{q})}{|\mathbf{q}|},
\label{HTL-energy-scalar}
\eeq
by doing the integration over $q$, we obtain
\beq
m_{s,{\rm LO}}^2={\cal P}_s=\frac{1}{2}\lambda T^2,
\label{mass-ms}
\eeq
which is half of $m^2_{d,{\rm LO}}$.
Its difference with SYM$_{4,4}$ is shown in Tab.~\ref{scalar-self-energy-weights}.
\begin{table}[t]
\centering
\caption{Relative weights of the one-loop scalar self-energy in
SYM$_{2,4}$ compared with SYM$_{4,4}$.}
\label{scalar-self-energy-weights}
\begingroup
\setlength{\tabcolsep}{8pt}
\renewcommand{\arraystretch}{1.5}
\def\diagcell#1{%
\raisebox{-0.3\height}{%
\includegraphics[width=0.06\textwidth]{scalar_#1.pdf}%
}%
}
\begin{tabular}{c|c}
\hline
Self-energy diagram
&
$\mathcal{P}_s^{\mathcal{N}=2}/
\mathcal{P}_s^{\mathcal{N}=4}$
\\[2pt]
\hline
\diagcell{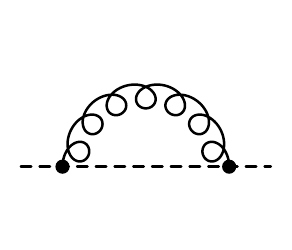}\quad
\diagcell{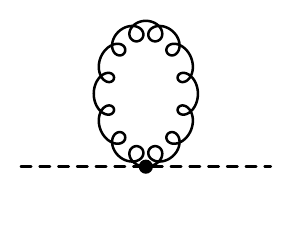}
& $1$
\\[4pt]
\diagcell{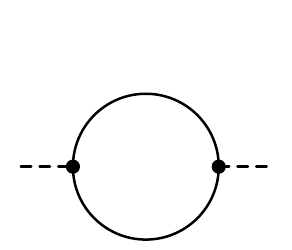}
& $\displaystyle\frac{1}{2}$
\\[4pt]
\diagcell{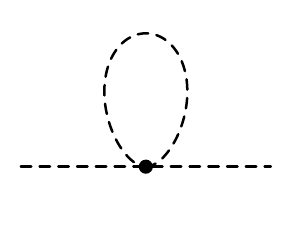}
& $\displaystyle\frac{1}{5}$
\\[4pt]
\hline
\end{tabular}%
\endgroup
\end{table}
The HTL fermionic self-energy is 
\beq
\Sigma_{f,ab}^{ij}(p)=
\frac{1}{2}\lambda\delta_{ab}\delta^{ij}
\int\frac{q^2\,dq}{2\pi^2}
\frac{f(\mathbf{q})}{|\mathbf{q}|}
\int\frac{d\Omega}{4\pi}
\frac{\slashed{y}}{p\cdot y} ,
\label{HTL-energy-fermion}
\eeq
so that, the fermion mass is 
\beq
m_{q,{\rm LO}}^2=\frac{1}{2}\lambda
\int\frac{q^2\,dq}{2\pi^2}
\frac{f(\mathbf{q})}{|\mathbf{q}|}=\frac{1}{4}\lambda T^2,
\label{mass-mq}
\eeq
which is one quarter of $m^2_{d,{\rm LO}}$.
Its difference with SYM$_{4,4}$ is shown in Tab.~\ref{fermion-self-energy-weights}.
\begin{table}[t]
\centering
\caption{Relative weights of the one-loop fermionic self-energy in
SYM$_{2,4}$ compared with SYM$_{4,4}$.}
\label{fermion-self-energy-weights}
\begingroup
\setlength{\tabcolsep}{10pt}
\renewcommand{\arraystretch}{1.5}
\def\diagcell#1{%
\raisebox{-0.5\height}{%
\includegraphics[width=0.09\textwidth]{fermion_diagram_#1.pdf}%
}%
}
\def\weightfont{\fontsize{8}{8}\selectfont}
\begin{tabular}{c|c}
\hline
Self-energy diagram
&
$\Sigma_f^{\mathcal{N}=2}/\Sigma_f^{\mathcal{N}=4}$
\\[4pt]
\hline
\diagcell{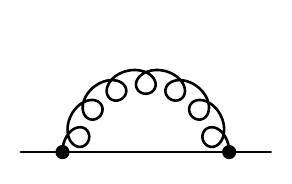}
& {\weightfont $1$}
\\[5pt]
\diagcell{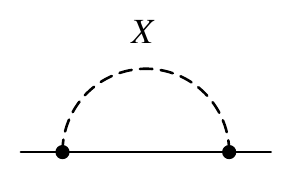}\quad
\diagcell{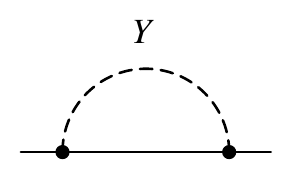}
& {\weightfont $\displaystyle\frac{1}{3}$}
\\[5pt]
\hline
\end{tabular}%
\endgroup
\end{table}
These three mass parameters can also be obtained by using the full
self-energies without invoking the HTL approximation. As a double check, we derive them in App.~\ref{full-self-energy}.

\section{Next-to-leading order thermodynamic potential}
\label{sec:potential}

The HTL perturbative thermodynamic potential at next-to-leading order in ${\cal N}=2$ SYM can be expressed as 
\beq
 \Omega_{\rm NLO}
 =\Omega_{\rm LO}+\Omega_{\rm 2-loop}
 +\Omega_{\rm HTL}+\Delta\Omega_{\rm NLO}.
\label{eq:omega-nlo}
\eeq
The four terms on the right-hand side are the leading order thermodynamic potential $\Omega_{\rm LO}$, the two-loop HTL free energy $\Omega_{\rm 2-loop}$, the two-loop HTL counterterm $\Omega_{\rm HTL}$, and the
renormalization counterterms $\Delta\Omega_{\rm NLO}$, respectively.

In this section, we present the explicit expressions of each term in Eq.~(\ref{eq:omega-nlo}) and discuss how they differ from ${\rm SYM}_{4,4}$. For convenience, we normalize our results to the ideal free energy,
\beq
 {\cal F}_{\rm ideal}^{\mathcal N=2}
 =-\frac{\dA\pi^2T^4}{12},
\label{eq:ideal}
\eeq
with $\dA=\Nc^2-1$ being the dimension of the adjoint representation. We also define dimensionless variables,
\begin{equation}
\begin{aligned}
 \hat{m}_d&=\frac{m_d}{2\pi T},&
 \hat{m}_s&=\frac{m_s}{2\pi T},\\
 \hat{m}_q&=\frac{m_q}{2\pi T},&
 \hat{\mu}&=\frac{\mu}{2\pi T}.
\end{aligned}
\label{eq:hatted}
\end{equation}

\subsection{Leading order thermodynamic potential}
\label{sec:one-loop}

\begin{figure}[htbp]
    \centering      
    \includegraphics[width=0.43\textwidth]{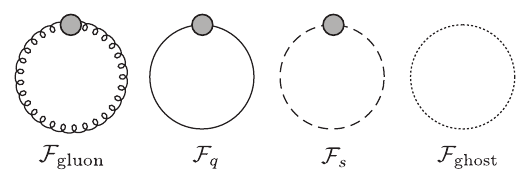} 
    \caption{One loop Feynman diagrams for ${\cal N}=2$ SYM theory in HTLpt. Shaded circles indicate HTL-dressed propagators.  }
    \label{1-loop}
\end{figure}

The leading order contribution includes the 1-loop diagrams shown in Fig.~\ref{1-loop} and the LO vacuum renormalization counterterm, 
\beq
 \Omega_{\rm LO}
 =\dA{\cal F}_g+2\dA{\cal F}_q+2\dA{\cal F}_s
 +\Delta_0{\cal E}_0,
\label{eq:omega-lo-components}
\eeq
where ${\cal F}_g={\cal F}_{\rm gluon}+{\cal F}_{\rm ghost}$, ${\cal F}_q$, and ${\cal F}_s$ are the one-loop HTL
free energies of a gauge field, a Majorana fermion and a scalar, which are the same in ${\rm SYM}_{4,4}$. The coefficient ``2" represents the number of independent Majorana fermions and scalars, while it is ``4'' and ``6" in ${\rm SYM}_{4,4}$. 
By using the results of ${\cal F}_g$, ${\cal F}_q$, and ${\cal F}_s$ in Refs.~\cite{Du:2020odw,Du:2022uyw}, our HTL one-loop free energy can be expressed as 
\begin{equation}
\begin{aligned}
\frac{\Omega_{1\text{-loop}}}
     {\mathcal{F}_{\mathrm{ideal}}^{\mathcal{N}=2}}
=& 1 - 2\hat{m}_d^2 + 8\hat{m}_d^3- 4\hat{m}_s^2 + 16\hat{m}_s^3
- 8\hat{m}_q^2 \\&+ 16(6-\pi^2)\hat{m}_q^4+ \frac{3}{2}\hat{m}_d^4
\bigg[ \frac{1}{\epsilon} + 2\gamma_E + \frac{2\pi^2}{3}\\
&- \frac{16}{3} - \frac{8}{3}\log 2 + 2\log\frac{\hat{\mu}}{2}
\bigg]\\
&+ 6\hat{m}_s^4 \bigg[\frac{1}{\epsilon}+ 2\gamma_E + 2\log\frac{\hat{\mu}}{2} \bigg],
\end{aligned}
\label{eq:omega-one-loop}
\end{equation}
where $\gamma_E$ is the Euler–Mascheroni constant. Since the LO vacuum energy counterterm $\Delta_0{\cal E}_0$ should cancel the divergences in $\Omega_{1\text{-loop}}$, we obtain
\begin{align}
\Delta_0{\cal E}_0&=\mathcal{F}_{\mathrm{ideal}}^{\mathcal{N}=2} \bigg(-\frac{3}{2\epsilon}\hat{m}_{d}^4-\frac{6}{\epsilon}\hat{m}_s^4\bigg)\nonumber\\&=
d_A\bigg(\frac{1}{128\pi^2\epsilon}m_d^4+\frac{1}{32\pi^2\epsilon}m_s^4\bigg),
\label{eq:vacuum-0-ct}
\end{align}
where the subscript ``0'' on $\Delta$ indicates that $\Delta_0{\cal E}_0$ is the zeroth-order term in the expansion of the counterterm $\Delta{\cal E}_0$ about $\delta=0$.
According to the HTLpt prescription, 
\begin{equation}
g \to \sqrt{\delta}\,g,
\quad m_n^2 \to (1-\delta)m_n^2,\quad  n=d,s,q, 
\end{equation}
the counterterm $\Delta{\cal E}_0$ has the form 
\begin{align}
\Delta{\cal E}_0&=
\bigg(\frac{d_A}{128\pi^2\epsilon}+{\cal O}(\lambda\delta)\bigg)(1-\delta)^2m_d^4 \nonumber\\&
+\bigg(\frac{d_A}{32\pi^2\epsilon}+{\cal O}(\lambda\delta)\bigg)(1-\delta)^2m_s^4.
\label{eq:vacuum-ct}
\end{align}
By combining Eqs.~(\ref{eq:omega-one-loop}) and (\ref{eq:vacuum-0-ct}), the final LO result in the RDR scheme is
\begin{align}
\frac{\Omega_{\rm LO}}
     {\mathcal{F}_{\mathrm{ideal}}^{\mathcal{N}=2}}
=& 1 - 2\hat{m}_d^2 + 8\hat{m}_d^3- 4\hat{m}_s^2 + 16\hat{m}_s^3
- 8\hat{m}_q^2 \nonumber\\&+ 16(6-\pi^2)\hat{m}_q^4+ \frac{3}{2}\hat{m}_d^4
\bigg[  2\gamma_E + \frac{2\pi^2}{3}- \frac{16}{3}\nonumber\\
& - \frac{8}{3}\log 2 + 2\log\frac{\hat{\mu}}{2}
\bigg]+ 12\hat{m}_s^4 \bigg[ \gamma_E + \log\frac{\hat{\mu}}{2} \big].
\label{eq:omega-lo}
\end{align}
The $m_d^4$ finite term differs from the result in the DRG scheme because RDR keeps the four-dimensional gauge-field representation size integer~\cite{Du:2022uyw}.

\subsection{Two-loop contribution in the RDR scheme }
\label{sec:two-loop}

\begin{figure}[htbp]
    \centering      
    \includegraphics[width=0.43\textwidth]{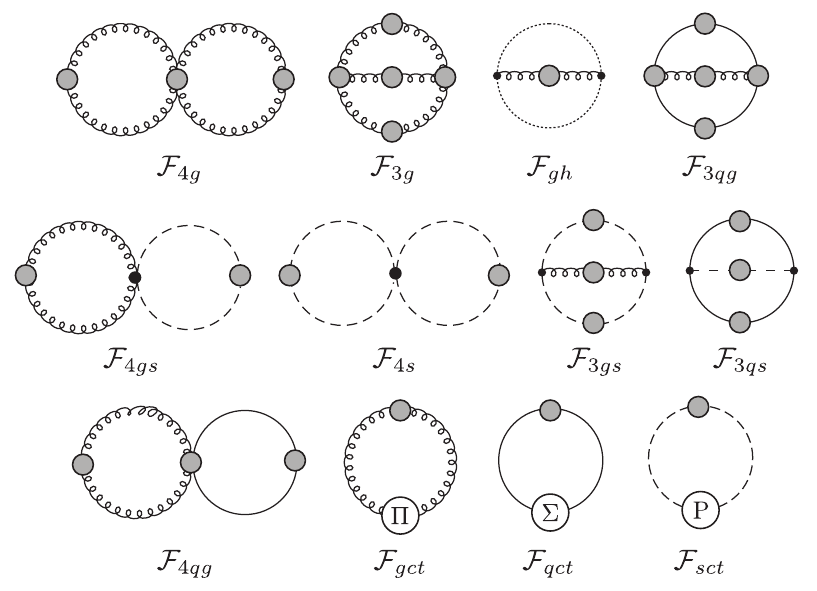} 
    \caption{Two loop Feynman diagrams for ${\cal N}=2$ SYM theory in HTLpt. Diagrams with the self-energies $\Pi$, ${\rm P}$ and $\Sigma$ contribute to HTL counterterms.}
    \label{2-loop}
\end{figure}

The two-loop contribution includes HTL two-loop free energy $\Omega_{2\text{-loop}}$ and counterterm $\Omega_{\rm HTL}$ which are shown in Fig.~\ref{2-loop}. The HTL two-loop free energy is expressed as
\begin{align}
\Omega_{2\text{-loop}}
= &d_A \lambda [
\mathcal{F}_{3g}+ \mathcal{F}_{4g}+ \mathcal{F}_{gh}
+ \mathcal{F}_{4s}+ \mathcal{F}_{3gs}+ \mathcal{F}_{4gs}\nonumber\\&
+ \mathcal{F}_{3qg}+ \mathcal{F}_{4qg}+ \mathcal{F}_{3qs}].
\end{align}
Comparing with ${\rm SYM}_{4,4}$ theory, results from diagrams with fermion-gluon-fermion vertex, scalar-gluon-scalar vertex, scalar-gluon-scalar-gluon vertex, four scalar vertex, and fermion-scalar-fermion vertex change. By using the relation in Eq.~(\ref{alpha-beta-matrix}), we obtained the relation of the HTL two-loop free energy between ${\rm SYM}_{2,4}$ and ${\rm SYM}_{4,4}$ which is shown in Tab.~\ref{tab:weights}.
As a result, the HTL two-loop free energy of ${\rm SYM}_{2,4}$ theory can be obtained from ${\rm SYM}_{4,4}$ theory,
\begin{align}
\Omega_{2\text{-loop}}=&d_A\lambda[{\cal F}_{3g+4g+gh}^{{\rm SYM}_{4,4}}
+\frac{1}{3}{\cal F}_{4s+3gs+4gs}^{{\rm SYM}_{4,4}} 
+\frac{1}{2}{\cal F}_{3qg+4qg}^{{\rm SYM}_{4,4}}\nonumber\\&
+\frac{1}{6} {\cal F}_{3qs}^{{\rm SYM}_{4,4}}
-2 {\cal F}_{\rm ext}],
\label{2-loop-N2}
\end{align}
where the term $-2{\cal F}_{\rm ext}$ is used to obtain the correct form of ${\cal F}_{4s}$, since the coefficient multiplying ${\cal F}_{4s}^{{\rm SYM}_{4,4}}$ is ``1/3'' not ``1/15'', and
\beq
 {\cal F}_{\rm ext}\equiv\left[\sumint_Q\frac{1}{Q^2 + m_s^2}\right]^2.
\label{eq:quartic-Q}
\eeq
\begin{table}[htbp]
\caption{Relative weights of the HTL two-loop free energy of ${\rm SYM}_{2,4}$ to that of ${\rm SYM}_{4,4}$ in Fig.~\ref{2-loop}.}
\label{tab:weights}
\begin{ruledtabular}
\begin{tabular}{lc}
Diagram & ${\cal F}^{{\cal N}=2}/{\cal F}^{{\cal N}=4}$ \\
\hline
${\cal F}_{3g}$, ${\cal F}_{4g}$, ${\cal F}_{gh}$ & $1$ \\
${\cal F}_{4s}$ & $1/15$ \\
${\cal F}_{3gs}$, ${\cal F}_{4gs}$ & $1/3$ \\
${\cal F}_{3qg}$, ${\cal F}_{4qg}$ & $1/2$ \\
${\cal F}_{3qs}$ & $1/6$ \\
\end{tabular}
\end{ruledtabular}
\end{table}
By expanding Eq.~(\ref{eq:quartic-Q}) in powers of $m_s/T$ and truncating at order $(m_s/T)^3$ to include all terms through $\lambda^{5/2}$, we obtain its contributions from the hard-hard (hh), hard-soft (hs) and soft-soft (ss) regions, respectively.
\begin{align}
{\cal F}_{\rm ext}^{hh}&=\sumint_{QP}\frac{1}{P^2 Q^2}-2m_s^2\sumint_{QP}\frac{1}{P^4 Q^2}, \nonumber\\
{\cal F}_{\rm ext}^{hs}&=2T\int_{\bold{p}} \frac{1}{p^2+m_s^2}\sumint_{Q}
\left(\frac{1}{Q^2} -\frac{m_s^2}{Q^4}\right),\nonumber\\
{\cal F}_{\rm ext}^{ss}&=T^2\int_{\bold{pq}} \frac{1}{(q^2+m_s^2)(p^2+m_s^2)} ,
\label{Fext}
\end{align}
with 
\begin{align}
&\sumint_Q \frac{1}{Q^2}
= \frac{T^2}{12}\left(\frac{\mu}{4\pi T}\right)^{2\epsilon}
\bigg[1+2\bigg(1+\frac{\zeta'(-1)}{\zeta(-1)}\bigg)\epsilon\bigg],\nonumber\\
&\sumint_Q \frac{1}{Q^4}
= \frac{1}{(4\pi)^2}
\left(\frac{\mu}{4\pi T}\right)^{2\epsilon}
\left[
\frac{1}{\epsilon}
+2\gamma_E
+\left(\frac{\pi^2}{4}-4\gamma_1\right)\epsilon
\right],\nonumber\\
&\int_{\mathbf{q}} \frac{1}{q^2+m_s^2}
= -\frac{m_s}{4\pi}
\left(\frac{\mu}{2m_s}\right)^{2\epsilon}
\left[1+2\epsilon\right],
\label{inte-2}
\end{align}
where $\gamma_1$ is the first Stieltjes constant. By using Eq.~(\ref{inte-2}), Eq.~(\ref{Fext}) reduces to
\begin{align}
{\cal F}_{\rm ext}^{hh}&=\frac{T^4}{144}
-\frac{m_s^2 T^2}{96\pi^2}
\left(
\frac{1}{\epsilon}+2\gamma_E-24\zeta'(-1)+2
\right)
\left(\frac{\mu}{4\pi T}\right)^{4\epsilon},\nonumber \\
{\cal F}_{\rm ext}^{hs}&=-\frac{m_s T^3}{24\pi}
+\frac{m_s^3T}{32\pi^3}
\left(\frac{1}{\epsilon}+2+2\gamma_E\right)
\left(\frac{\mu}{2m_s}\right)^{2\epsilon}
\left(\frac{\mu}{4\pi T}\right)^{2\epsilon}
,\nonumber\\
{\cal F}_{\rm ext}^{ss}&=\frac{m_s^2 T^2}{16\pi^2} .
\label{Fextfinal}
\end{align}
By using Eqs.~(\ref{2-loop-N2}), (\ref{Fextfinal}) and results in Refs.~\cite{Du:2020odw,Du:2022uyw}, our HTL two-loop free energy can be expressed as 
\begin{widetext}
\begin{equation}
\begin{aligned}
\frac{\Omega_{2\text{-loop}}}     {\mathcal{F}_{\mathrm{ideal}}^{\mathcal{N}=2}}
&=\frac{\lambda}{\pi^2} \biggl\{
-\frac{3}{4}+3\hat{m}_s+3\hat{m}_d-3\hat{m}_s\hat{m}_d
+3\hat{m}_s^2\hat{m}_d
-12\hat{m}_d\hat{m}_q^2
+\hat{m}_d^2\big(1.245328+3\log\hat{m}_d\big)
\\&-\frac{3}{2}\hat{m}_d^2\hat{m}_s
\bigg(\frac{1}{\epsilon}+2+2\gamma_E-2\log\hat{m}_s+4\log\frac{\hat{\mu}}{2}
\bigg)-3\hat{m}_q^2\bigg(\frac{1}{\epsilon}+9.967511+4\log\frac{\hat{\mu}}{2}\bigg)+12\hat{m}_q^2\hat{m}_s\\& \times
\bigg(\frac{1}{\epsilon}+2+2\gamma_E+4\log 2-2\log\hat{m}_s+4\log\frac{\hat{\mu}}{2}
\bigg)+\hat{m}_s^2\big(9.1182-2\gamma_E+24\zeta'(-1)+6\log\hat{m}_s
\big)
\\&+\frac{3}{2}\hat{m}_d^3
\bigg(\frac{1}{\epsilon}+2-\frac{8}{3}\log 2+2\gamma_E-2\log\hat{m}_d+4\log\frac{\hat{\mu}}{2}\bigg)-\hat{m}_s^3(1+12\log 2)\biggr\}, 
\end{aligned}
\label{omega-two-loop}
\end{equation}
\end{widetext}
which is gauge invariant. 

The HTL two-loop counterterm $\Omega_{\rm HTL}$ is expressed as 
\beq
\Omega_{\rm HTL}=d_A{\cal F}_{gct}+2d_A{\cal F}_{qct}+2d_A{\cal F}_{sct}. 
\label{FHTL}
\eeq
As shown in Eqs.~\eqref{HTL-energy-gluon}, \eqref{HTL-energy-scalar}, and \eqref{HTL-energy-fermion}, the gluon, scalar, and fermion HTL self-energies in SYM$_{2,4}$ and SYM$_{4,4}$ differ only in their mass parameters. Consequently, the corresponding counterterm contributions $\mathcal{F}_{gct}$, $\mathcal{F}_{sct}$, and $\mathcal{F}_{qct}$
have the same functional forms, with the appropriate mass
parameters for each theory.
As a result, by using the results in Refs.~~\cite{Du:2020odw,Du:2022uyw}, our HTL two-loop counterterm is obtained 
\begin{equation}
\begin{aligned}
\frac{\Omega_{\rm HTL}}{\mathcal{F}_{\mathrm{ideal}}^{\mathcal{N}=2}}
&=4\hat{m}_s^2-24\hat{m}_s^3+2\hat{m}_d^2-12\hat{m}_d^3+8\hat{m}_q^2
\\
&+32\hat{m}_q^4(\pi^2-6)-12\hat{m}_s^4\left(\frac{1}{\epsilon}+2\gamma_E+2\log\frac{\hat{\mu}}{2}\right)
\\
&-3\hat{m}_d^4\left(\frac{1}{\epsilon}-\frac{16}{3}+2\gamma_E+\frac{2}{3}\pi^2-\frac{8}{3}\log 2+2\log\frac{\hat{\mu}}{2}
\right).
\label{omega-HTL}
\end{aligned}
\end{equation} 
Moreover, it can also be calculated by substituting $m_d^2\rightarrow(1-\delta)m_d^2$, $m_s^2\rightarrow(1-\delta)m_s^2$  and $m_q^2\rightarrow(1-\delta)m_q^2$ in the one-loop expression of ${\cal F}_g$, ${\cal F}_s$, and ${\cal F}_q$ and expanding them to linear order in $\delta$. The linear order with $\delta=1$ is the result of $\Omega_{\rm HTL}$.

\subsection{Renormalization and final NLO result}
\label{sec:renormalization}

The ultraviolet divergences in Eqs.~(\ref{omega-two-loop}) and (\ref{omega-HTL}) should be removed by the renormalization of the HTL mass parameters $m_d$, $m_s$ and $m_q$, and the renormalization of the vacuum energy density ${\cal E}_0$, respectively.
For the former case, the divergences are
\begin{align}
\Omega_{\rm 2-loop}^{\rm div}
&={\cal F}_{\rm ideal}^{\mathcal N=2}\frac{\lambda}{\pi^2\epsilon}\left[
-3\hat{m}_q^2
+\frac{3}{2}\hat{m}_d^3
-\frac{3}{2}\hat{m}_s\hat{m}_d^2
+12\hat{m}_s\hat{m}_q^2
\right]\nonumber\\
&=\frac{d_A\lambda}{\epsilon}\bigg(\frac{m_q^2 T^2}{16\pi^2}
-\frac{m_d^3 T}{64\pi^3}+\frac{m_s m_d^2 T}{64\pi^3}
-\frac{m_s m_q^2 T}{8\pi^3} \bigg).
\label{eq:remaining-pole}
\end{align}
In order to cancel Eq.~(\ref{eq:remaining-pole}), we need to construct a combination of HTL masses $m_i^2$ and $\partial \Omega_{\rm LO}/\partial m_i^2$. We define matrices ${\rm M}\equiv (m_d^2,m_s^2,m_q^2)$, ${\rm B}\equiv(\frac{\partial\Omega_{\rm LO}}{\partial m_d^2},\frac{\partial\Omega_{\rm LO}}{\partial m_s^2},\frac{\partial\Omega_{\rm LO}}{\partial m_q^2})$, which are connected by a $3\times3$ matrix ${\rm A}$, and satisfy
\beq
({\rm M}_n {\rm A}_{nr} {\rm B}_r+\Omega_{\rm 2-loop}^{\rm div})|_{\rm div}=0,
\label{condition}
\eeq
where $n,r=1,2,3$. Regarding Eq.~(\ref{condition}), we need to mention that: (1) the overall factor of $\lambda$ in Eq.~(\ref{eq:remaining-pole}) implies that the leading-order result is needed only through order $m_d^3$ and $m_s^3$ to obtain contributions through order $\lambda^{5/2}$; (2) the overall factor of $1/\epsilon$ implies that we must retain terms proportional of order $\epsilon$ in Eq.~(\ref{eq:omega-lo}) to get the correct result of ${\rm M}_n {\rm A}_{nr} {\rm B}_r$.
Since only terms proportional to $m_q^2$, $m_d^3$, $m_s m_d^2$ and $m_s m_q^2$ exist, other terms must vanish, we obtain 
\begin{equation}
{\rm A}=\begin{pmatrix}
-\dfrac{\lambda}{8\pi^2\epsilon}
& \dfrac{\lambda}{16\pi^2\epsilon}& 0 \\[6pt]
0 & 0 & 0 \\[6pt]
0& -\dfrac{\lambda}{2\pi^2\epsilon}
& -\dfrac{\lambda}{8\pi^2\epsilon}
\end{pmatrix},
\label{matrix-A}
\end{equation}
which is non-diagonal, similarly to SYM$_{4,4}$.
As a result, ${\rm M}_n {\rm A}_{nr} {\rm B}_r$ can be written as 
\beq
\Delta_1 m_d^2\frac{\partial\Omega_{\rm LO}}{\partial m_d^2} +\left(
4\Delta_1 m_q^2-\frac{1}{2}\Delta_1 m_d^2
\right)\frac{\partial\Omega_{\rm LO}}{\partial m_s^2}
+\Delta_1 m_q^2\frac{\partial\Omega_{\rm LO}}{\partial m_q^2}, 
\label{mass-counterterm}
\eeq
with
\beq
\Delta_1 m_d^2=-\frac{\lambda}{8\pi^2\epsilon}m_d^2,
\quad\quad
\Delta_1 m_q^2=-\frac{\lambda}{8\pi^2\epsilon}m_q^2.
\label{mass-counterterms}
\eeq
where the subscript ``1'' means $\Delta_1 m_d^2$ and $\Delta_1 m_q^2$ are the first-order terms in the expansion of the counterterms $\Delta m_d^2$ and $\Delta m_q^2$ about $\delta=0$, and then set $\delta=1$.
The coefficient of $\frac{\partial\Omega_{\rm LO}}{\partial m_s^2}$ in Eq.~(\ref{mass-counterterm}) is irrelevant to $\Delta_1 m_s^2$ and is predictable, since there is no divergence proportional to $m_s^2$ in Eq.~(\ref{eq:remaining-pole}), otherwise, it would introduce an uncanceled pole proportional to $m_s^2$.
According to the HTLpt prescription, we obtain
\begin{align}
\Delta m_d^2={}&\bigg(-\frac{\delta\lambda}{8\pi^2\epsilon}+{\cal O}(\lambda^2\delta^2)\bigg)(1-\delta)m_d^2,
\nonumber\\
\Delta m_q^2={}&\bigg(-\frac{\delta\lambda}{8\pi^2\epsilon}+{\cal O}(\lambda^2\delta^2)\bigg)(1-\delta)m_q^2.
\label{eq:mass-counterterms}
\end{align}
By using Eq.~(\ref{mass-counterterm}), the renormalization counterterms $\Delta\Omega_{\rm NLO}$ can be written as 
\begin{align}
\Delta\Omega_{\rm NLO}&=
\Delta_1{\cal E}_0
+\Delta_1m_d^2\frac{\partial\Omega_{\rm LO}}{\partial m_d^2}+\Delta_1m_q^2\frac{\partial\Omega_{\rm LO}}{\partial m_q^2}\nonumber\\
&
+\left(
4\Delta_1 m_q^2-\frac{1}{2}\Delta_1 m_d^2
\right)\frac{\partial\Omega_{\rm LO}}{\partial m_s^2},
\label{eq:nlo-renormalization}
\end{align}
where $\Delta_1{\cal E}_0$ is obtained by expanding
Eq.~\eqref{eq:vacuum-ct} through first order in $\delta$, and setting $\delta=1$. As a result, 
\begin{align}
\Delta_1\mathcal{E}_0&= d_A \bigg(-\frac{1}{64\pi^2\epsilon}m_d^4-\frac{1}{16\pi^2\epsilon}m_s^4 \bigg)\nonumber\\&
={\cal F}_{\rm ideal}^{\mathcal N=2}\left(
\frac{3}{\epsilon}\hat{m}_d^4 +\frac{12}{\epsilon}\hat{m}_s^4
\right), 
\end{align}
which cancels the divergences in Eq.~(\ref{omega-HTL}) of $\Omega_{\rm HTL}$.
Therefore, the result of $\Delta\Omega_{\rm NLO}$ is obtained, 
\begin{equation}
\begin{aligned}
\frac{\Delta\Omega_{\mathrm{NLO}}}{{\cal F}_{\rm ideal}^{\mathcal N=2}}&=
\frac{3\hat{m}_d^4}{\epsilon}+\frac{12\hat{m}_s^4}{\epsilon}
+\frac{\lambda}{\pi^2}\Biggl[\left(-12\hat{m}_s\hat{m}_q^2
+\frac{3}{2}\hat{m}_s\hat{m}_d^2\right)\\& \times
\left(\frac{1}{\epsilon}+2-2\log\hat{m}_s+2\log\frac{\hat{\mu}}{2}
\right)\\&
-\frac{3}{2}\hat{m}_d^3\left(\frac{1}{\epsilon}+2-2\log\hat{m}_d
+2\log\frac{\hat{\mu}}{2}\right)
\\&
+3\hat{m}_q^2\left(\frac{1}{\epsilon}-24\zeta'(-1)-\frac{2}{3}\log 2+2+2\log\frac{\hat{\mu}}{2}\right)\Biggr].
\label{HTL-renormaliza}
\end{aligned}
\end{equation}
We need to emphasize that there is no coupling counterterm, since it first contributes at order $\delta^2$ and hence does not enter $\Omega_{\rm NLO}$, which has also been referred to in QCD \cite{Andersen:2002ey} and SYM$_{4,4}$ \cite{Du:2020odw}. 

By combining Eqs.~(\ref{eq:omega-lo}), (\ref{omega-two-loop}), (\ref{omega-HTL}) and (\ref{HTL-renormaliza}), the renormalized NLO thermodynamic potential can be written as
\begin{widetext}
\begin{align}
\frac{\Omega_{\rm NLO}}{{\cal F}_{\rm ideal}^{\mathcal N=2}}
=& 1-8\hat{m}_s^3-4\hat{m}_d^3
+16\hat{m}_{q}^4(-6+\pi^2)
+\hat{m}_s^4\bigg(-12\gamma_E-12\log\frac{\hat{\mu}}{2}\bigg)
+\hat{m}_d^4\bigg(8-3\gamma_E-\pi^2+4\log2-3\log\frac{\hat{\mu}}{2}\bigg)\nonumber\\
&+\frac{\lambda}{\pi^2}\Bigg[
-\frac{3}{4}+3\hat{m}_s+3\hat{m}_d-3\hat{m}_s\hat{m}_d
-12\hat{m}_d \hat{m}_{q}^2
-\hat{m}_s^3\left(1+12\log2\right)
+\hat{m}_s^2\bigg(5.148092-2\gamma_E +3\hat{m}_d\nonumber\\
&+6\log\hat{m}_s\bigg)
-3\hat{m}_s\hat{m}_d^2\bigg(\gamma_E+\log\frac{\hat{\mu}}{2}\bigg)
+24\hat{m}_s \hat{m}_{q}^2\bigg(\gamma_E+2\log2+\log\frac{\hat{\mu}}{2}\bigg)
+\hat{m}_d^2(1.245328+3\log\hat{m}_d) \nonumber\\&
+\hat{m}_{q}^2\left(-11.992211-2\log2-6\log\frac{\hat{\mu}}{2}\right)
+\hat{m}_d^3\left(3\gamma_E-4\log2+3\log\frac{\hat{\mu}}{2}\right)
\Bigg].
\label{eq:omega-nlo-final}
\end{align}
\end{widetext}
Equation~\eqref{eq:omega-nlo-final} is finite and gauge independent. To check our result against the known result through order ${\cal O}(\lambda^{3/2})$, we insert Eq~(\ref{eq:hatted}) and the leading-order expressions of masses (\ref{mass-md}), (\ref{mass-ms}) and  (\ref{mass-mq}) into Eq.~(\ref{eq:omega-nlo-final}), and expand through order $\lambda^2$,
\begin{align}
\frac{\Omega_{\rm NLO}}{{\cal F}_{\rm ideal}^{\mathcal N=2}}\bigg|_{\rm strict}&= 1-\frac{3\lambda}{4\pi^2}+\frac{\lambda^{3/2}}{\pi^3}
\left(1+\frac{1}{\sqrt{2}}\right)
+\frac{\lambda^2}{\pi^4}
\bigg(\frac{3}{4}\log\lambda \nonumber\\&
-\frac{3}{4}\log\frac{\hat{\mu}}{2}
-3.130102-\frac{5}{8}\gamma_E
\bigg) +{\cal O} (\lambda^{5/2}),
\label{F-static}
\end{align}
where terms through order $\lambda^{3/2}$ agree exactly with the known results~\cite{Vazquez-Mozo:1999yck,Paik:2009iz,Tantary:2026udn}, and the coefficient of the term $\lambda^2\log \lambda$ also agrees with the strict form~\cite{Tantary:2026udn}.
The remaining terms at order $\lambda^2$ are not complete, and a three-loop HTL calculation is required. In particular, the coefficient of $\lambda^2 \log {\hat{\mu}}$ for the strict result in Ref.~\cite{Tantary:2026udn} is $-3/16$, rather than $-3/4$, because the three-loop HTLpt diagrams and the coupling-renormalization counterterm first enter at next-to-next-to-leading order.

\section{Gap equations}
\label{sec:numerics}

The gluon, scalar and quark mass parameters $m_d$, $m_s$ and $m_q$ are determined by using the variational method,
\begin{align}
&\frac{\partial}{\partial m_q} \Omega_{\rm NLO}(T,\lambda,m_d,m_s,m_q,\delta=1)=0,\nonumber\\
&\frac{\partial}{\partial m_d} \Omega_{\rm NLO}(T,\lambda,m_d,m_s,m_q,\delta=1)=0,\nonumber\\
&\frac{\partial}{\partial m_s} \Omega_{\rm NLO}(T,\lambda,m_d,m_s,m_q,\delta=1)=0,
\label{eq:gap-equations}
\end{align}
which can be reduced to,
\begin{equation}
\begin{aligned}
\hat{m}_q^2(\pi^2-6)
={}&\frac{\lambda}{8\pi^2}
\Biggl[3\hat{m}_d-6\hat{m}_s
\left(\gamma_E+2\log 2+\log\frac{\hat{\mu}}{2}
\right)
\\
&\qquad
+\frac{3}{2}\log\frac{\hat{\mu}}{2}+3.344626\Biggr],
\label{eq:gap-mq}
\end{aligned}
\end{equation}
and 
\begin{equation}
\begin{aligned}
&\hat{m}_d^2-\frac{1}{3}\hat{m}_d^3\left(8-3\gamma_E-\pi^2
+4\log 2-3\log\frac{\hat{\mu}}{2}
\right)
\\
&=\frac{\lambda}{4\pi^2}\Biggl[1-\hat{m}_s+\hat{m}_d
\left(1+\frac{2}{3}\left(1.245328+3\log\hat{m}_d\right)
\right)
\\
&\qquad
-2\hat{m}_d\hat{m}_s\left(\gamma_E+\log\frac{\hat{\mu}}{2}
\right)+\hat{m}_s^2-4\hat{m}_q^2
\\
&\qquad
+\hat{m}_d^2\left(3\gamma_E-4\log 2+3\log\frac{\hat{\mu}}{2}\right)\Biggr],
\label{eq:gap-md}
\end{aligned}
\end{equation}
and 
\begin{equation}
\begin{aligned}
&\hat{m}_s^2+2\hat{m}_s^3\left(\gamma_E+\log\frac{\hat{\mu}}{2}\right)
\\
&=\frac{\lambda}{8\pi^2}\Biggl[1-\hat{m}_d
-\frac{1}{3}\hat{m}_s\left(-16.296185
+4\gamma_E-12\log\hat{m}_s\right)
\\
&\qquad+2\hat{m}_s\hat{m}_d-\hat{m}_s^2(1+12\log 2)
-\hat{m}_d^2\left(\gamma_E+\log\frac{\hat{\mu}}{2}
\right)
\\
&\qquad
+8\hat{m}_q^2\left(\gamma_E+2\log 2+\log\frac{\hat{\mu}}{2}
\right)\Biggr],
\end{aligned}
\label{eq:gap-ms}
\end{equation}
respectively. 
Note that terms proportional to $\hat{m}_q^2$ in Eqs.~(\ref{eq:gap-md}) and (\ref{eq:gap-ms}) can be written in terms of $\hat{m}_d$ and $\hat{m}_s$ by using Eq.~(\ref{eq:gap-mq}).
We solve Eqs.~\eqref{eq:gap-md}--\eqref{eq:gap-ms} numerically on the positive stationary branch that is continuously connected to the perturbative thermal masses of $m_d$ and $m_s$ after inserting Eq.~(\ref{eq:gap-mq}) into Eqs.~(\ref{eq:gap-md}) and (\ref{eq:gap-ms}). 
Since the coupling $\lambda$ is a function of the renormalization scale $\mu$, and satisfies Eq.~(\ref{eq:beta}), by imposing the boundary condition
$\lambda(\mu_0)$,
we separate variables and integrate to obtain
\begin{equation}
\int_{\lambda(\mu_0)}^{\lambda(\mu)}
\frac{d\lambda'}{\lambda'^2}=-\frac{1}{4\pi^2}
\int_{\mu_0}^{\mu}\frac{d\mu'}{\mu'}.
\end{equation}
Choosing the reference scale $\mu_0=2\pi T$, we obtain \footnote{Unless otherwise specified, $\lambda$ denotes $\lambda(\hat{\mu})$ throughout this paper.}
\begin{equation}
\lambda(\hat{\mu})=\frac{\lambda(2\pi T)}{
1+\dfrac{\lambda(2\pi T)}{4\pi^2}\log\hat{\mu}}.
\label{lam-run}
\end{equation}

\begin{figure}[htbp]
    \centering      \includegraphics[width=0.38\textwidth]{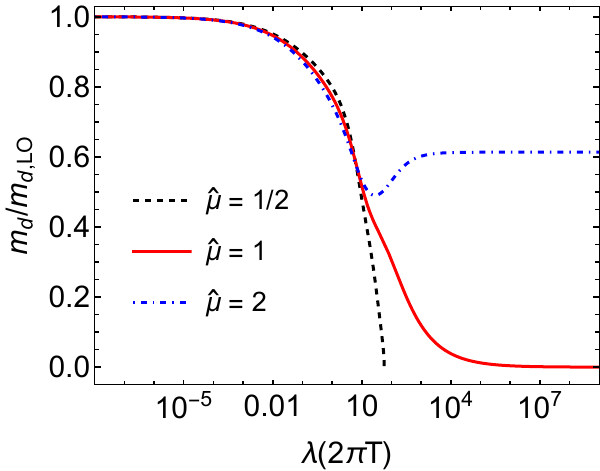}
\includegraphics[width=0.38\textwidth]{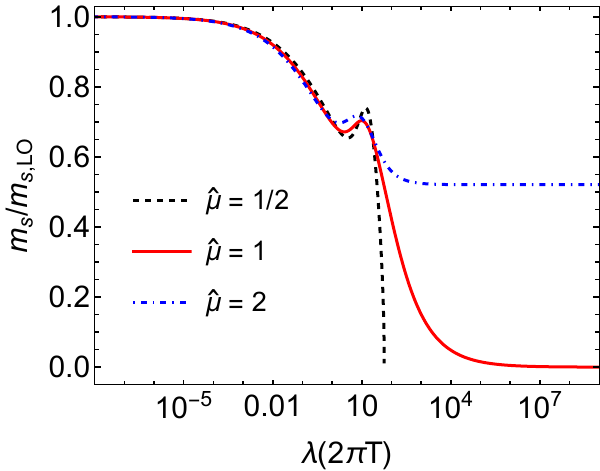} 
\includegraphics[width=0.38\textwidth]{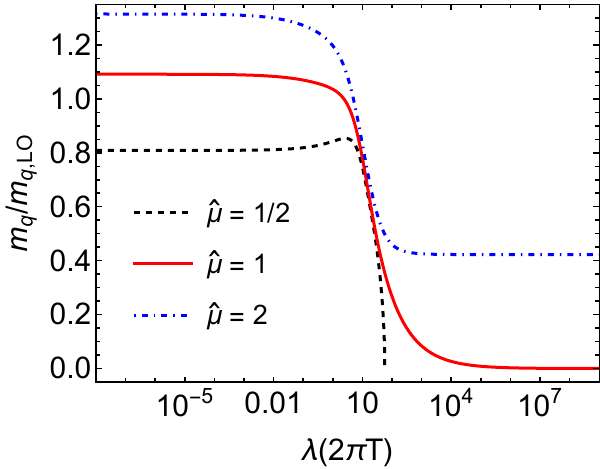}
    \caption{Numerical solutions of the gap equations for $m_d$, $m_s$, and $m_q$ as a function of $\lambda(2\pi T)$.  }
    \label{numerical-mass}
\end{figure}
Inserting Eq.~(\ref{lam-run}) into gap equations (\ref{eq:gap-md}), (\ref{eq:gap-ms}) and (\ref{eq:gap-mq}), we get the numerical solutions for scaled $m_d$, $m_s$, and $m_q$ as a function of $\lambda(2\pi T)$ shown in Fig.~\ref{numerical-mass}. In all three panels, the red line is the solution when taking $\hat{\mu}=1$, the black dashed line is $\hat{\mu}=1/2$, and the blue dash-dotted line is $\hat{\mu}=2$. As expected, in the weak-coupling limit $\lambda(2\pi T)\rightarrow 0^{+}$, all mass-dependent terms inside the square brackets on the right-hand side of Eqs.~(\ref{eq:gap-md}) and (\ref{eq:gap-ms}) vanish, while the higher-order mass terms on the left-hand side are negligible compared with the quadratic term, which yields 
\begin{align}
&\hat{m}_d^2|_{\lambda(2\pi T)\rightarrow 0}\rightarrow \frac{\lambda}{4\pi^2}=\hat{m}_{d,{\rm LO}}^2,\nonumber \\  &\hat{m}_s^2|_{\lambda(2\pi T)\rightarrow 0}\rightarrow\frac{\lambda}{8\pi^2}=\hat{m}_{s,{\rm LO}}^2, 
\label{lamto0}
\end{align}
and is reflected in Fig.~\ref{numerical-mass}, where the normalized ratio approaches unity. Putting Eq.~(\ref{lamto0}) into (\ref{eq:gap-mq}), we obtain
\begin{equation}
\hat{m}_q^2|_{\lambda(2\pi T)\rightarrow 0}\rightarrow\frac{\lambda}
{16\pi^2(\pi^2-6)}\left(6.689253+3\log\frac{\hat{\mu}}{2}\right),
\end{equation}
which depends on $\hat{\mu}$. Since 
\beq
\hat{m}_{q,{\rm LO}}^2=\frac{\lambda}{16\pi^2}, 
\label{lamto01}
\eeq
the mass ratio $m_q/m_{q,{\rm LO}}|_{\hat{\mu}=1}\rightarrow 1.091461$, $m_q/m_{q,{\rm LO}}|_{\hat{\mu}=1/2}\rightarrow 0.808646$ and $m_q/m_{q,{\rm LO}}|_{\hat{\mu}=2}\rightarrow 1.31479$, rather than approaching unity.
The same behaviors in the perturbative limit also appear in the $\mathcal N=4$ NLO calculation~\cite{Du:2020odw} and QCD~\cite{Andersen:2003zk}. As $\lambda(2\pi T)$ increases, the scaled $m_d$ and $m_q$ decrease; the scaled $m_s$ exhibits nonmonotonic behavior, with a brief increase interrupting its overall decrease. The three curves of $\hat{\mu}$ converge around $\lambda(2\pi T)\simeq 10$ before separating again.
At very large coupling, as a formal limit of the truncated gap equations, all three ratios are equal to zero for $\hat\mu=1$, and equal to positive finite constants for $\hat\mu=2$. We analytically proved this in App.~\ref{app:mass-in-infinity} at $\lambda(2\pi T)\to\infty$. 
For $\hat{\mu}=1/2$, since the pole of coupling in Eq.~(\ref{lam-run}) is $\lambda(2\pi T)=4\pi^2/\log2$, these mass ratios will terminate at this point, with the ratios tending to zero.

\begin{figure}[htbp]
    \centering      
    \includegraphics[width=0.38\textwidth]{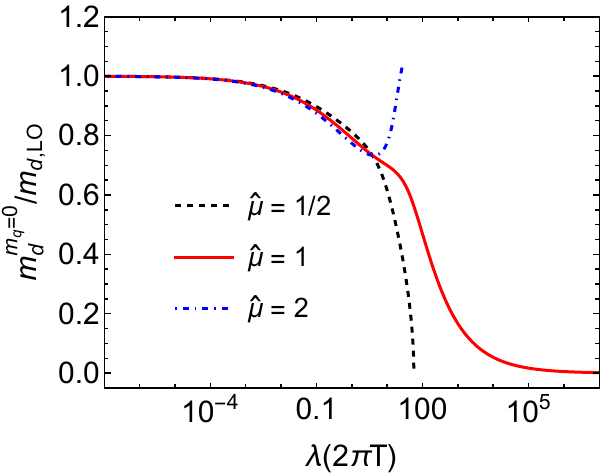} 
    \includegraphics[width=0.38\textwidth]{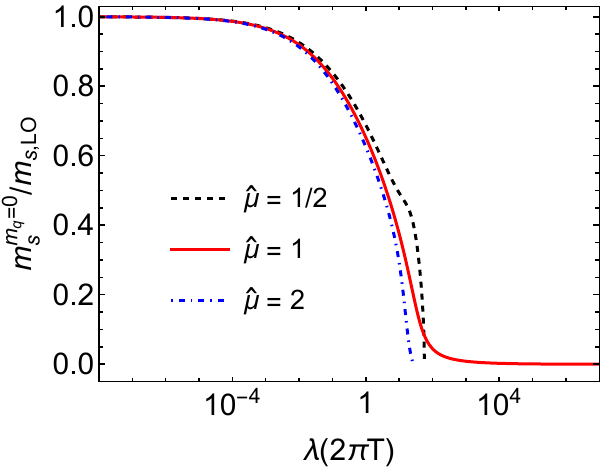} 
    \caption{Numerical solutions of gap equations for scaled $m_d$ and $m_s$ as a function of $\lambda(2\pi T)$ at $m_q=0$.  }
    \label{mass-mq0}
\end{figure}

\begin{figure}[htbp]
    \centering      
    \includegraphics[width=0.38\textwidth]{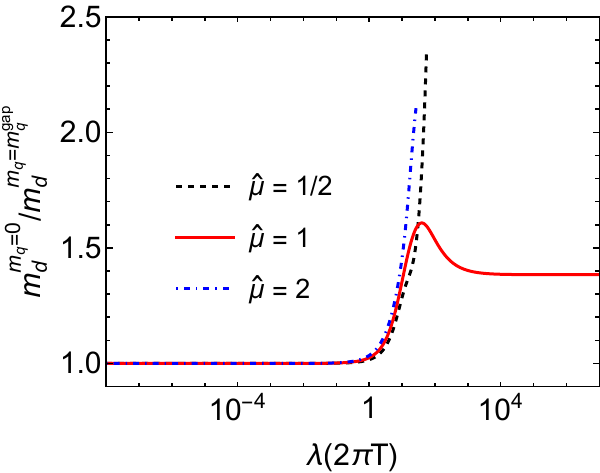}  
    \includegraphics[width=0.38\textwidth]{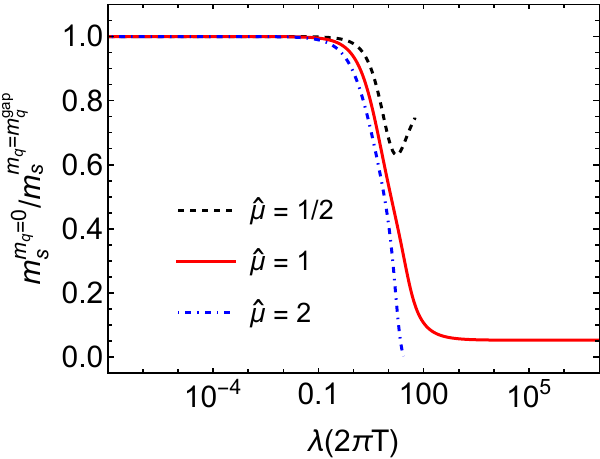} 
    \caption{Ratio of the numerical solution of $m_d$ and $m_s$ for the two quark mass schemes: $m_q=0$ and $m_q=m_q^{\rm gap}$.  }
    \label{mass-ratio}
\end{figure}

On the other hand, the variational $m_q$ does not reproduce the correct leading-order perturbative coefficient as $\lambda(2\pi T)\rightarrow0$, and increasing the order of the HTLpt calculation may not resolve this issue due to the fact that the fermionic sector is infrared safe. We consider an alternative reorganization by setting $m_q=0$ for the HTL effective Lagrangian, as has been done in SYM$_{4,4}$, to provide a qualitative assessment of the sensitivity of $m_d$ and $m_s$ to the $m_q$ prescription\footnote{ Several other prescriptions can be adopted for $m_q$, such as setting $m_q=m_{q,{\rm LO}}$ for all values of $\lambda$; taking $m_q=1/2 m_d$ or $1/\sqrt{2}m_s$, where the coefficient is determined by the ratio of their leading-order values of Eqs.~(\ref{mass-md}), (\ref{mass-ms}) and (\ref{mass-mq}). Here we consider $m_q=0$ just to assess the sensitivity of $m_d$, $m_s$ and the thermodynamics to it, and compare with SYM$_{4,4}$.} Fig.~\ref{mass-mq0} shows the numerical results of $m_d$ and $m_s$ at $m_q=0$. Compared with Fig.~\ref{numerical-mass}, it indicates that setting $m_q=0$ will not qualitatively alter the behaviors for $\hat\mu=1$ and $\hat\mu=1/2$, but changes the behavior for $\hat\mu=2$. In the latter case, the positive mass branch terminates at $\lambda(2\pi T)\simeq27.31$, with $m_d/m_{d,\mathrm{LO}}\simeq1.038 $ and $m_s/m_{s,\mathrm{LO}}\sim 10^{-5}$, indicating a sensitivity of the gap masses to the prescription.
To quantify this sensitivity, we present the ratios of the solutions obtained with $m_q=0$ to those obtained with $m_q=m_q^{\mathrm{gap}}$ (\ref{eq:gap-mq}) in Fig.~\ref{mass-ratio}. It shows that setting $m_q=0$ increases $m_d$ and decreases $m_s$ relative to the full variational prescription.
As $\lambda(2\pi T)$ increases, the mass ratio $(m_d^{m_q=0}/m_d^{m_q=m_q^{\mathrm{gap}}},m_s^{m_q=0}/m_s^{m_q=m_q^{\mathrm{gap}}})$ for $\hat\mu=1$ eventually approaches a constant value of $(1.38399,0.05267)$ by using Eqs.~(\ref{mass-infinity-u1}) and (\ref{mass-infinity-u1-mq0}); for $\hat\mu=2$, it terminates at $(2.11688,10^{-5})$  with $\lambda(2\pi T)\simeq27.31$; for $\hat\mu=1/2$, it terminates at $(2.3971,0.746749)$  as $\lambda(2\pi T)$ approaches the running-coupling pole $4\pi^2/\log2$. These behaviors suggest that the gluon and scalar gap masses exhibit a strong dependence on the fermion mass prescription at large coupling.
 
\section{Thermodynamic functions} 
\label{numerics-themodynamics}
\begin{figure*}[t]
    \centering      
\includegraphics[width=0.38\textwidth]{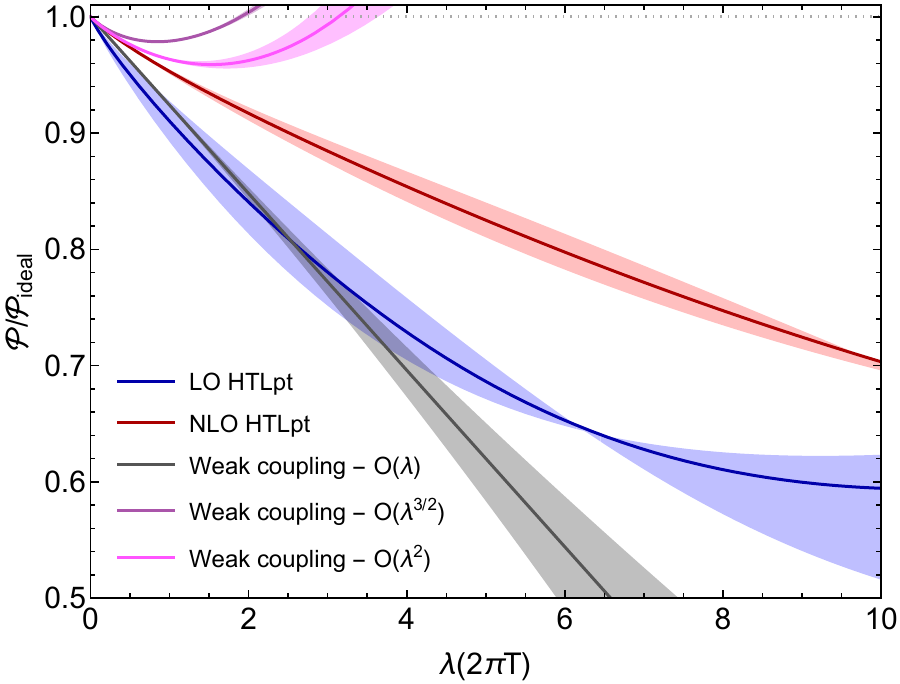} \qquad\qquad
\includegraphics[width=0.38\textwidth]{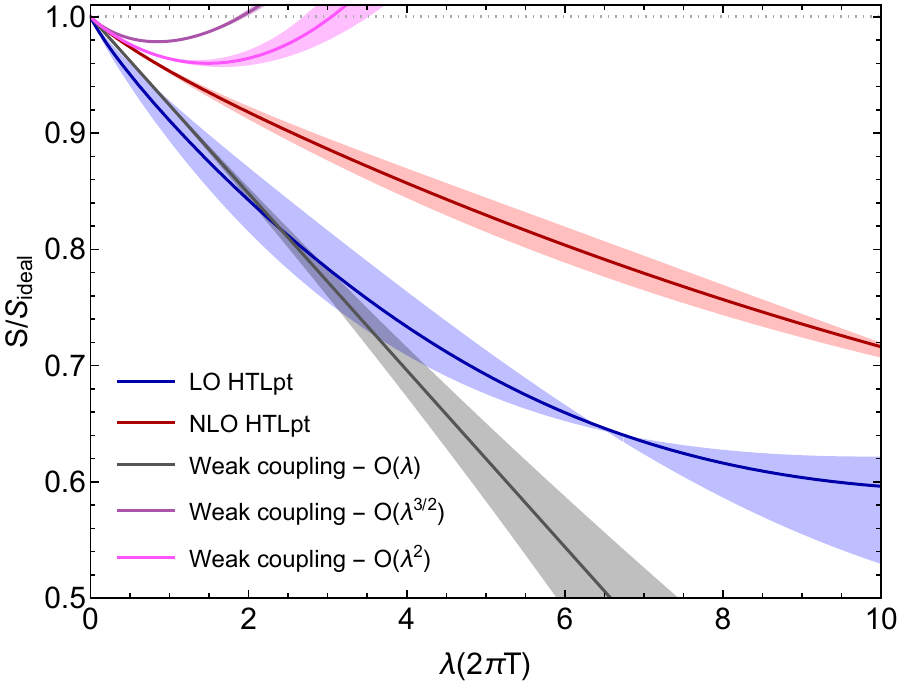} 
\par\medskip
\includegraphics[width=0.38\textwidth]{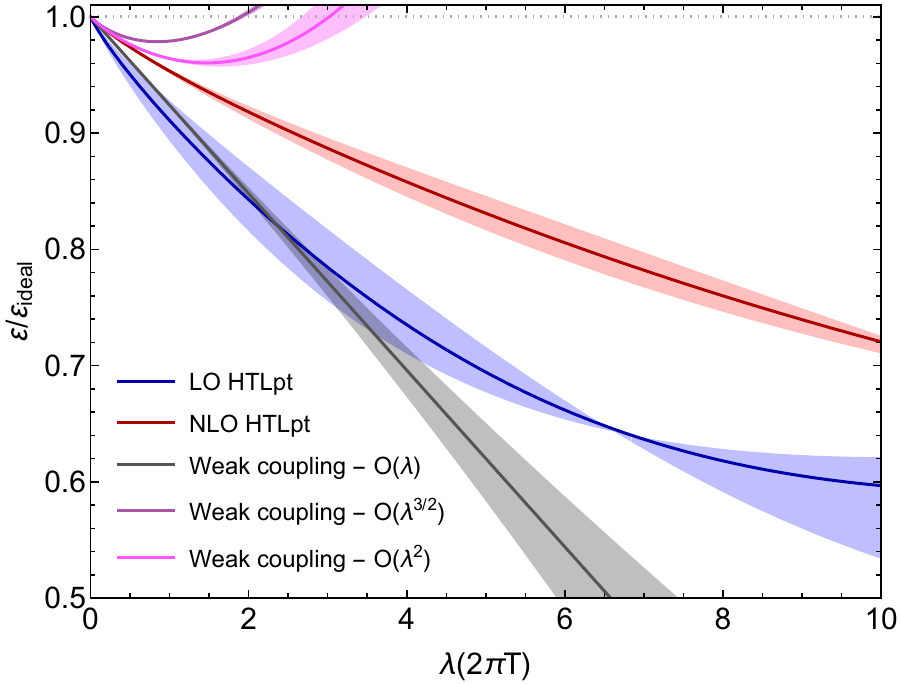} \qquad\qquad
\includegraphics[width=0.38\textwidth]{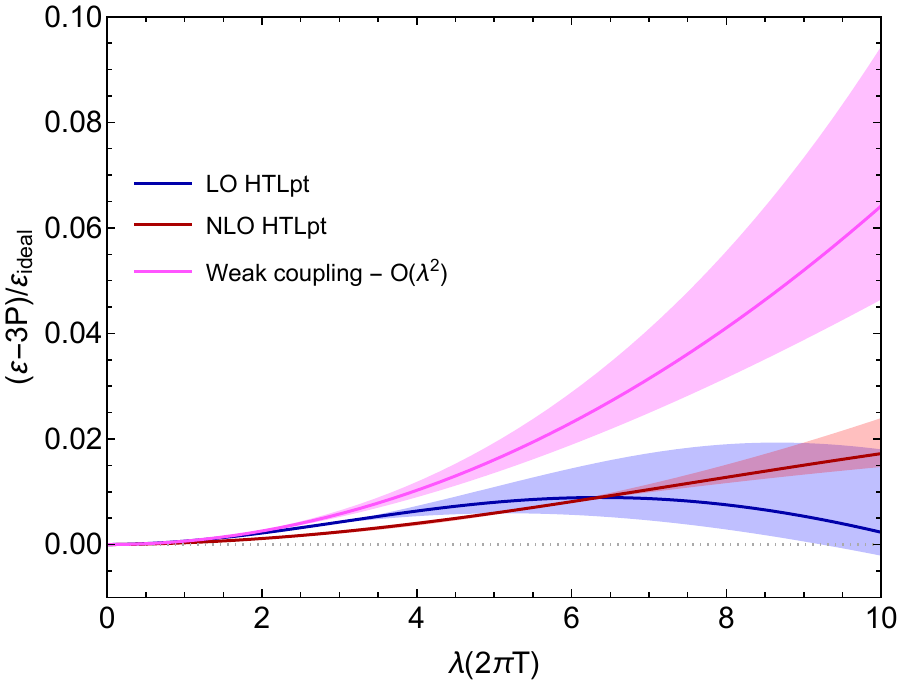} 
    \caption{Comparison of LO and NLO HTLpt results for scaled pressure, entropy, energy density, and trace anomaly with the known weak coupling result up to order ${\cal O}(\lambda^2)$~\cite{Tantary:2026udn}. The shaded band corresponds to varying the renormalization scale $\mu$ by a factor of two around $\mu=2\pi T$. } 
    \label{thermodynamics}
\end{figure*}

\begin{figure*}[t]
    \centering      
\includegraphics[width=0.38\textwidth]{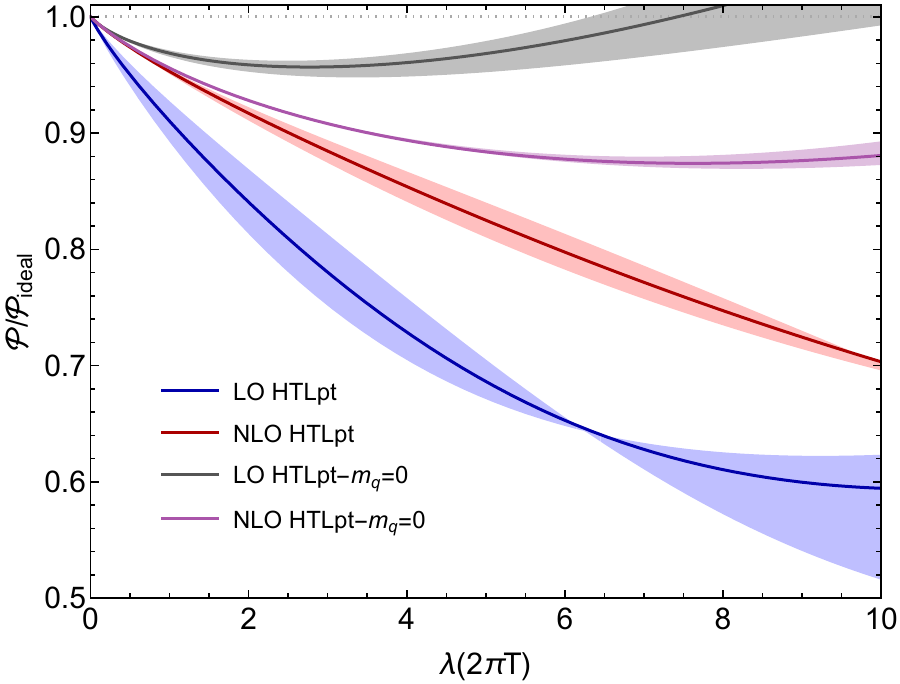} \qquad\qquad
\includegraphics[width=0.38\textwidth]{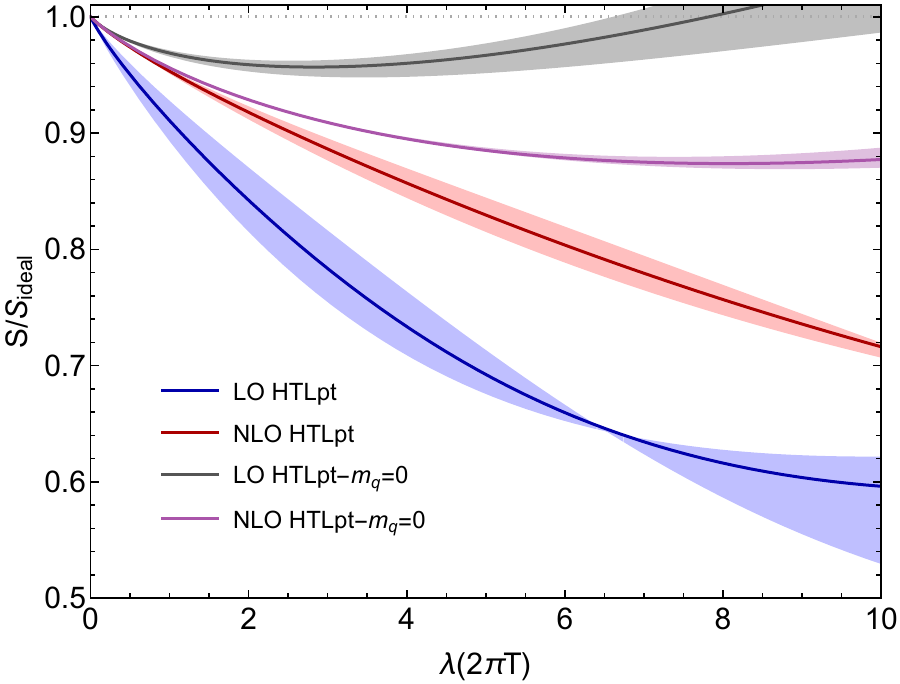} 
\par\medskip
\includegraphics[width=0.38\textwidth]{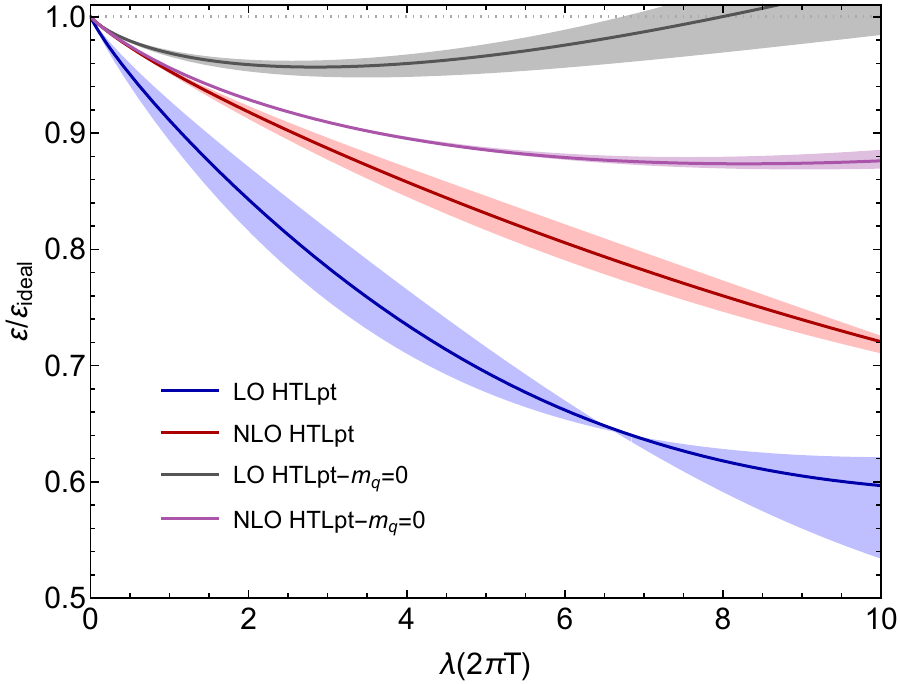} \qquad\qquad
\includegraphics[width=0.38\textwidth]{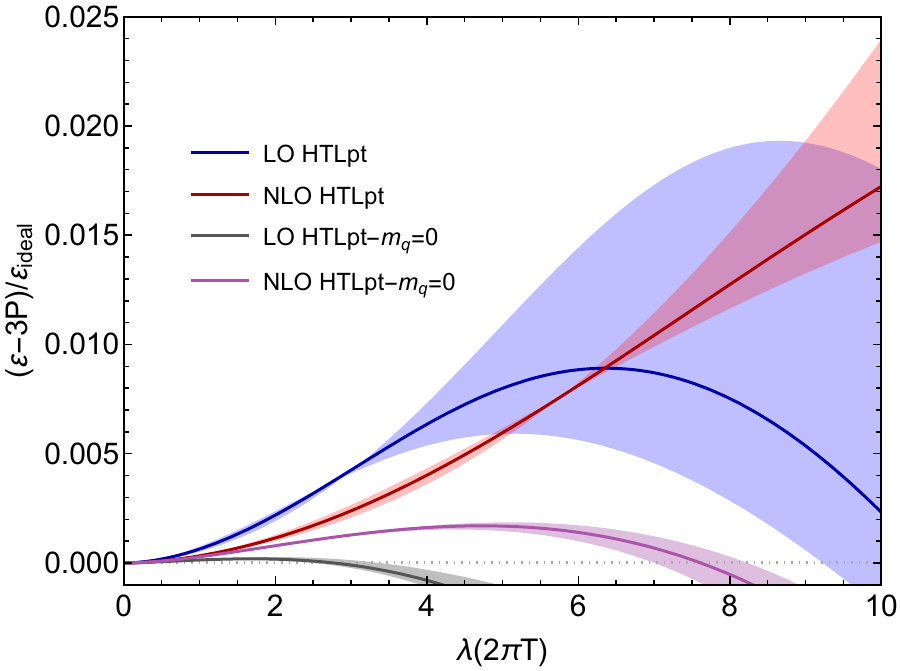} 
    \caption{Comparison of LO and NLO HTLpt results for scaled pressure, entropy, energy density, and trace anomaly with the corresponding results by setting $m_q=0$.} 
    \label{thermodynamics-1}
\end{figure*}

In this section, we investigate the thermodynamics of $\mathcal{N}=2$ SYM, and examine the effects of different fermion mass prescriptions. The NLO HTLpt approximation to the free energy is obtained by evaluating the NLO HTLpt thermodynamic potential (\ref{eq:omega-nlo-final}) at the solution of the gap equation (\ref{eq:gap-equations}),
\beq
\mathcal{F}_{\mathrm{NLO}}=\Omega_{\mathrm{NLO}}\left(
T,\lambda, m_d^{\mathrm{gap}},m_s^{\mathrm{gap}},m_q^{\mathrm{gap}},\delta=1\right).
\eeq
The pressure, entropy density, energy density and the trace anomaly can then be obtained by using
\begin{align}
&\mathscr{P} = -\mathcal{F},\quad
\mathcal{S} = -\frac{d\mathcal{F}}{dT},\quad
\mathcal{E} = \mathcal{F}
-T\frac{d\mathcal{F}}{dT},\nonumber\\
&\mathcal{E}-3\mathscr{P}=-T^5\frac{d}{dT}
\left(\frac{\mathcal{F}}{T^4}\right).
\label{themornamics}
\end{align} 
Unlike ${\cal N}=4$ SYM, ${\cal N}=2$ SYM is not conformal and has a running coupling. Consequently, the pressure, entropy density, and energy density, normalized to their ideal limits, 
\begin{align}
&\mathscr{P}_{\rm ideal}=-{\cal F}_{\rm ideal}^{\mathcal N=2}=\frac{\dA\pi^2T^4}{12},\nonumber\\
&\mathcal{S}_{\rm ideal} = \frac{\dA\pi^2T^3}{3},\quad \mathcal{E}_{\rm ideal} = \frac{\dA\pi^2T^4}{4},
\label{themornamics-ideal}
\end{align}
from Eq.~(\ref{eq:ideal}), do not generally coincide. By using Eqs.~(\ref{themornamics}), (\ref{themornamics-ideal}) and (\ref{eq:beta}) with $\mu=2\pi T\hat{\mu}$ at fixed $\hat{\mu}$, and using the chain rule, we obtain their relations,
\begin{align}
&\frac{{\cal S}}{{\cal S}_{\rm ideal}}=\frac{\mathscr{P}}{\mathscr{P}_{\rm ideal}}-\frac{\lambda^2}{16\pi^2}\frac{d}{d\lambda}\bigg(\frac{\mathscr{P}}{\mathscr{P}_{\rm ideal}}\bigg),\nonumber\\
&\frac{{\cal E}}{{\cal E}_{\rm ideal}}=\frac{\mathscr{P}}{\mathscr{P}_{\rm ideal}}-\frac{\lambda^2}{12\pi^2}\frac{d}{d\lambda}\bigg(\frac{\mathscr{P}}{\mathscr{P}_{\rm ideal}}\bigg),
\label{eq:thermodynamic-ratios-1}
\end{align}
and 
\beq
\frac{{\cal E}-3\mathscr{P}}{{\cal E}_{\rm ideal}}=-\frac{\lambda^2}{12\pi^2}\frac{d}{d \lambda}\bigg(\frac{\mathscr{P}}{\mathscr{P}_{\rm ideal}}\bigg).
\label{eq:thermodynamic-ratios-2}
\eeq
If the coupling $\lambda$ does not run, Eq.~(\ref{eq:thermodynamic-ratios-2}) and the second terms in Eq.~(\ref{eq:thermodynamic-ratios-1}) vanish, which return back to the result of ${\cal N}=4$ SYM.
By using Eq.~(\ref{F-static}), we obtain scaled pressure, entropy and energy density in the static perturbative limit up to order $\lambda^2$,
\begin{align}
&\frac{\mathcal{S}}{\mathcal{S}_{\mathrm{ideal}}}\bigg|_{\rm strict}
= 1-\frac{3\lambda}{4\pi^2}
+\frac{\lambda^{3/2}}{\pi^3}\bigg(1+\frac{1}{\sqrt{2}}\bigg)+\frac{\lambda^2}{\pi^4}\bigg(\frac{3}{4}\log\lambda
\nonumber\\
&\qquad -\frac{3}{4}\log\frac{\hat\mu}{2}-3.083227-\frac{5\gamma_E}{8}
\bigg)+{\cal O} (\lambda^{5/2}),
\nonumber\\
&\frac{\mathcal{E}}{\mathcal{E}_{\mathrm{ideal}}}\bigg|_{\rm strict}
= 1-\frac{3\lambda}{4\pi^2}+\frac{\lambda^{3/2}}{\pi^3}\bigg(1+\frac{1}{\sqrt{2}}\bigg)+\frac{\lambda^2}{\pi^4}\bigg(\frac{3}{4}\log\lambda
\nonumber\\
&\qquad-\frac{3}{4}\log\frac{\hat\mu}{2}-3.067602-\frac{5\gamma_E}{8}
\bigg)+{\cal O} (\lambda^{5/2}),
\end{align} 
with $\mathscr{P}/\mathscr{P}_{\mathrm{ideal}}|_{\rm strict}=  \Omega_{\rm NLO}/{\cal F}_{\rm ideal}^{\mathcal N=2}|_{\rm strict}
$. Their difference is only reflected in the constant coefficient of the term $\lambda^2$. Comparing with those obtained from Ref.~\cite{Tantary:2026udn}, we find agreement at order $\lambda^2$ only in the coefficient of the term $\lambda^2\ln\lambda$. 
Our scaled trace anomaly in the static perturbative limit is 
\beq
\frac{{\cal E}-3\mathscr{P}}{{\cal E}_{\rm ideal}}\bigg|_{\rm strict}
= \frac{\lambda^2}{16\pi^4}+{\cal O} (\lambda^{5/2}),
\eeq
which is the same as the result from Ref.~\cite{Tantary:2026udn}.

Fig.~\ref{thermodynamics} shows our HTLpt results for the scaled pressure, entropy, energy density, and trace anomaly in ${\cal N}=2$ SYM as a function of $\lambda(2\pi T)$. The red solid line with a red shaded band is the NLO HTLpt result, and the blue solid line with a blue shaded band is the LO HTLpt result determined by evaluating Eqs.~(\ref{eq:thermodynamic-ratios-1}) and (\ref{eq:thermodynamic-ratios-2}) at the solution to the NLO mass gap equations (\ref{eq:gap-equations}) with the solution of $\lambda$ in Eq.~(\ref{lam-run}). The HTLpt shaded bands result from the renormalization scale $\hat{\mu}=(1/2,1,2)$, with the central value of the renormalization scale plotted as solid blue and red lines for the LO and NLO results, respectively. 
The black solid line with a black shaded band, purple solid line with a purple shaded band and magenta solid line with a magenta shaded band are the weak-coupling result~\cite{Tantary:2026udn} truncated at order $\lambda$, $\lambda^{3/2}$ and $\lambda^2$, respectively. The gray dotted lines indicate the weak
coupling limits of ``1'' and ``0''.
In addition, we restrict our discussion to the regime $0\leq\lambda(2\pi T)\leq10$, where the upper limit is a practical choice because of the pronounced scale and mass-prescription dependence of the gap solutions at larger couplings.

As can be seen from Fig.~\ref{thermodynamics}, the LO and NLO HTLpt predictions for the normalized pressure, entropy density, and energy density approach unity in the weak-coupling limit, consistent with the preceding analytic analysis. As $\lambda(2\pi T)$ increases, the three quantities decrease smoothly throughout the range shown. At small coupling, the LO HTLpt results remain close to the weak-coupling predictions truncated at $\mathcal O(\lambda)$, although with the NLO gap masses the LO coefficient of $\lambda$ differs from the strict one. Their subsequent separation indicates that the additional contributions retained in HTLpt expressions become increasingly important.
The additional contributions retained in NLO HTLpt beyond $\mathcal O(\lambda^{3/2})$ lower the predictions truncated at that order and remove the upturn over the finite-coupling range.
Since NLO HTLpt does not include the complete $\mathcal O(\lambda^2)$ contribution, this difference reflects both the incomplete matching of the coefficients at that order and the higher-order contributions retained in the unexpanded expressions. 
The NNLO HTLpt result from the three-loop calculation would allow for a systematic comparison with the weak-coupling result through $\mathcal{O}(\lambda^2)$ even $\mathcal{O}(\lambda^{5/2})$ in the strict perturbative limit.
In addition, the NLO HTLpt results generally exhibit narrower scale-variation bands than the LO results at larger couplings, indicating reduced renormalization-scale dependence. Meanwhile, NLO HTLpt corrections are positive and become more pronounced with increasing coupling, reaching approximately 23\% around $\lambda(2\pi T)\simeq7\text{–}8$ relative to the LO results. These sizable corrections indicate that higher-order effects remain important in this regime.

On the other hand, the scaled trace anomaly shown in the lower-right panel of Fig.~\ref{thermodynamics} exhibits a different order dependence from the other three thermodynamic quantities. Both the LO and NLO HTLpt results vanish in the weak-coupling limit consistent with the preceding analytic analysis. The LO HTLpt central curve reaches a maximum around $\lambda(2\pi T)\sim 6$ and subsequently decreases, whereas the NLO HTLpt central curve increases monotonically over the range shown. 
At $\lambda(2\pi T)=10$, the normalized trace anomaly is approximately 0.017 at NLO, compared with approximately 0.065 for the weak-coupling result through $\mathcal O(\lambda^2)$. Thus, the NLO HTLpt prediction lies substantially below the corresponding weak-coupling prediction at larger couplings. Higher-loop calculations are needed to assess the stability of the HTLpt result. 
Moreover, the weak-coupling result exhibits a substantially broader scale-variation band than NLO HTLpt. 
These observations indicate that, despite the similar overall trends of the pressure, entropy density, and energy density at LO and NLO HTLpt, the trace anomaly remains sensitive to the order of the HTLpt approximation.

Fig.~\ref{thermodynamics-1} shows the comparison of the thermodynamics obtained by determining $m_q$ from the gap equation with those obtained by setting $m_q=0$. The blue, red and gray curves are the same as in Fig.~\ref{thermodynamics}, while the black and purple curves show the corresponding LO and NLO HTLpt results for $m_q=0$. In the weak-coupling limit, the normalized pressure, entropy density, and energy density approach unity in both prescriptions, while the trace anomaly vanishes. 
For the first three quantities, setting $m_q=0$ yields larger values than $m_q^{\rm gap}$ at the corresponding order, and modifies their coupling dependence as $\lambda(2\pi T)$ increases. In the case $m_q=0$, the LO curves exhibit nonmonotonic behavior with the minima around $\lambda(2\pi T)\simeq3$, and their values begin to be greater than one around $\lambda(2\pi T)\simeq 7.5$, while the NLO curves decrease more gradually and exhibit shallow minima around $\lambda(2\pi T)\simeq7\text{–}8$. 
The relative differences between the NLO results for the two prescriptions are maximal at $\lambda(2\pi T)=10$, reaching approximately $22\%\text{–}26\%$.
Moreover, the NLO HTLpt corrections exhibit opposite trends in the two prescriptions: they increase these three quantities with $m_q=m_q^{\rm gap}$, while suppressing them with $m_q=0$.
The scaled trace anomaly exhibits an even more pronounced sensitivity to the mass prescription. The NLO HTLpt central curve remains positive and increases monotonically with $m_q=m_q^{\rm gap}$; however, it is substantially suppressed when setting $m_q=0$, reaching a maximum around $\lambda(2\pi T)\simeq4.4\text{–}5.1$ before decreasing and becoming negative around $\lambda(2\pi T)\simeq7$. The LO trace anomaly in the latter case becomes negative at a smaller coupling.
Different from the previous results for SYM$_{4,4}$, our  SYM$_{2,4}$ results indicate that the quark mass prescription has a significant impact on the thermodynamics at the two-loop order.
A possible explanation is that the coupling runs and depends on $\hat{\mu}$ in $\mathcal N=2$ SYM whose solution is shown in Eq. (\ref{lam-run}), while it does not run in $\mathcal N=4$ SYM.

\section{Summary and Outlook}
\label{sec:summary}

In this paper, we have extended the LO and NLO HTLpt calculation of the thermodynamic potential in ${\cal N}=4$ SYM to the pure four-dimensional $\mathcal N=2$ SYM under the regularization
by the dimensional reduction scheme. By using relations of the Feynman rules between $\mathcal N=4$ and $\mathcal N=2$ SYM, we obtained the gluon, scalar, and fermion self-energies in SYM$_{2,4}$, including their general expressions and HTL limit, and then obtained their corresponding leading-order thermal masses. We obtained the LO and NLO HTLpt thermodynamic potential, and the NLO HTLpt result reproduces the strict free energy through order $\lambda^{3/2}$ and the coefficient of $\lambda^2\log\lambda$.
We also obtained the required vacuum-energy, gluon and quark mass counterterms, with no coupling and scalar mass counterterm entering at this order.

By using the gap equations together with the running coupling, we have numerically calculated the three thermal masses, and compared these solutions with those obtained by setting $m_q=0$. It indicates that both $m_d$ and $m_s$ exhibit non-negligible sensitivity to the fermionic thermal mass prescription. With these former thermal masses, we numerically calculated the scaled pressure, entropy density, energy density, and trace anomaly, and compared them with the known weak-coupling results in the static perturbative limit through $\mathcal O(\lambda^2)$. Our result shows that the LO and NLO HTLpt predictions of the first three quantities decrease monotonically with increasing $\lambda(2\pi T)$, and the NLO corrections raise the corresponding LO values. 
In addition, NLO HTLpt retains contributions beyond $\mathcal O(\lambda^{3/2})$, yielding a lower prediction and does not exhibit an upturn over the range. This suggests that the higher-loop calculation may improve the order-by-order stability of the thermodynamics, and extend their applicability to stronger couplings.      
The scaled trace anomaly displays a distinct coupling dependence, where the LO results exhibit nonmonotonic behavior whereas the NLO result remains positive and increases monotonically, which indicates that it remains sensitive to the order of the HTLpt expansion.
Moreover, the comparison with the $m_q=0$ prescription further demonstrates that the determination of $m_q$ affects not only the magnitudes of the thermodynamics but also the signs of the NLO corrections, where the NLO HTLpt results suppress the LO results for the first three quantities, and the trace anomaly changes from increasing to decreasing. 
The fact that the coupling runs and depends on $\hat{\mu}$ in ${\cal N}=2$ SYM may contribute to this $m_q$ prescription dependence.
In the future, we will extend our calculations to the three-loop HTLpt to examine the order-by-order stability of the thermodynamic predictions, and assess whether their sensitivity to the quark mass prescription is reduced at higher orders.

\begin{acknowledgments}
The work of Q.D. is supported by the National Natural Science Foundation of China with Grant No. 12305135, by the Guangxi Natural Science Foundation with Grant No. 2023GXNSFBA026027 
and by the Guangxi Young Elite Scientist Sponsorship Program with Grant No. GXYESS2026074; and U.T. acknowledges support from the Department of Mathematics and Natural Sciences at Prince Mohammad Bin Fahd University. We thank Jens O. Andersen and Michael Strickland for discussions of the earlier $\mathcal N=4$ HTLpt calculation.
\end{acknowledgments}

\appendix
\section{The one-loop self-energies beyond the HTL approximation }
\label{full-self-energy}
 
By using Tabs.~\ref{gluon-self-energy-weights}, \ref{scalar-self-energy-weights} and \ref{fermion-self-energy-weights},
the one-loop gluonic self-energy $\Pi^{ab}_{\mu\nu}(P)$ beyond the HTL approximation contains two parts,
\begin{align}
\Pi_{\mu\nu}^{b,ab}(P)
&= -\lambda\delta^{ab}
\bigg[
  \frac{D}{2}\bar{\Pi}_{\mu\nu}^{b}(P)
  - 2\bigl(P^2\delta_{\mu\nu}-P_\mu P_\nu\bigr) \nonumber\\&\times
  \sumint_Q \frac{1}{Q^2(P+Q)^2}
\bigg],
\label{gluon-self-1}
\\
\Pi_{\mu\nu}^{f,ab}(P)
&= \lambda\delta^{ab}
\bigg[
  2\bar{\Pi}_{\mu\nu}^{f}(P)
  - 2\bigl(P^2\delta_{\mu\nu}-P_\mu P_\nu\bigr)\nonumber\\&\times
  \sumint_{\{Q\}} \frac{1}{Q^2(P+Q)^2}
\bigg],
\label{gluon-self-2}
\end{align}
with 
\begin{align}
\bar{\Pi}_{\mu\nu}^{b}(P)
&\equiv 
\sumint_Q \frac{2\delta_{\mu\nu}}{Q^2}
- \sumint_Q
\frac{(2Q+P)_\mu(2Q+P)_\nu}{Q^2(P+Q)^2},
\label{gluon-self-3}
\\
\bar{\Pi}_{\mu\nu}^{f}(P)
& \equiv
\sumint_{\{Q\}} \frac{2\delta_{\mu\nu}}{Q^2}
- \sumint_{\{Q\}}
\frac{(2Q+P)_\mu(2Q+P)_\nu}{Q^2(P+Q)^2},
\label{gluon-self-4}
\end{align}
where 
\begin{align}
\sumint_{\{Q\}} \frac{1}{Q^2}
=&-\frac{T^2}{24}\left(\frac{\mu}{4\pi T}\right)^{2\epsilon}\nonumber\\ &\times\bigg[
1+2\bigg(1-\log 2+\frac{\zeta'(-1)}{\zeta(-1)}
\bigg)\epsilon\bigg],
\label{inte-1}
\end{align}
where $\{Q\}$ indicates that the sum runs over fermionic
Matsubara frequencies. The differences between ${\rm SYM}_{2,4}$ and ${\rm SYM}_{4,4}$ are the coefficients ``$\frac{D}{2}$" in Eq.~(\ref{gluon-self-1}) and the overall coefficient ``1'' in Eq.~(\ref{gluon-self-2}), which are ``$\frac{D+4}{2}$"  and  ``2'' in ${\rm SYM}_{4,4}$.  
By using the integrals in Eq.~(\ref{inte-2}) and Eq.~(\ref{inte-1}), the gluon mass is obtained 
\begin{align}
m_d^2&=-\Pi_{00}(0)=\lambda D(d-2)\sumint_{Q} \frac{1}{Q^2}\nonumber\\&-\lambda 4(d-2)\sumint_{\{Q\}} \frac{1}{Q^2}=\lambda T^2+{\cal O}(\epsilon),
\end{align}
by using the following identities
\begin{align}
 &\sumint_{Q} \frac{Q_0^2}{Q^4}=\frac{3-d}{2}\sumint_{Q} \frac{1}{Q^2},\nonumber\\
 &\sumint_{\{Q\}} \frac{Q_0^2}{Q^4}=\frac{3-d}{2}\sumint_{\{Q\}} \frac{1}{Q^2}, 
\end{align}
from integration by parts.
The scalar and fermionic self-energies are 
\begin{align}
{\cal P}_{s,ab}^{IJ}(P)&=\lambda \delta_{ab}\delta^{IJ}\bigg[ D \sumint_Q \frac{1}{Q^2}-4 \sumint_{\{Q\}} \frac{1}{Q^2} \nonumber\\ & -\sumint_Q\frac{2P^2}{Q^2(P+Q)^2}+\sumint_{\{Q\}}\frac{2P^2}{Q^2(P+Q)^2} \bigg], 
\label{scalar-self}\\
\Sigma_{f,ab}^{ij}(P)&=\lambda \delta_{ab}\delta^{ij}D\sumint_Q \frac{\slashed{P}-\slashed{Q}}{Q^2(P-Q)^2}.
\label{quark-self}
\end{align}
For ${\rm SYM}_{4,4}$, the coefficients of the first, second and fourth terms in square brackets in Eq.~(\ref{scalar-self}), and the overall coefficient of Eq.~(\ref{quark-self}) are ``$D+4$", ``$8$", ``$4$" and ``$D+4$", respectively. By using the integrals in Eq.~(\ref{inte-2}) and Eq.~(\ref{inte-1}), the scalar and fermion masses are obtained
\begin{align}
m_s^2={\cal P}_s(0)&=\lambda\bigg(D\sumint_{Q} \frac{1}{Q^2} -4\sumint_{\{Q\}} \frac{1}{Q^2}\bigg)\nonumber\\&=\frac{1}{2}\lambda T^2+{\cal O}(\epsilon),
\end{align} 
and 
\begin{align}
m_q^2&=\frac{D}{2}\lambda\bigg( \sumint_{Q} \frac{1}{Q^2} -\sumint_{\{Q\}} \frac{1}{Q^2}\bigg)\nonumber\\&=\frac{1}{4}\lambda T^2+{\cal O}(\epsilon).
\end{align} 
The higher-order terms in $\epsilon$ for $m_d^2$ and $m_s^2$ are used to calculate the free energy at order $\lambda^{5/2}$ \cite{Carrington:2026psm,Zhai:1995ac}.

\section{Proof of gluon and scalar mass ratios in the $\lambda(2\pi T) \to \infty$ limit for $\hat{\mu} = 1$ and $\hat{\mu} = 2$}
\label{app:mass-in-infinity}

From Eq.~(\ref{lam-run}), as $\lambda(2\pi T) \to \infty$, we obtain
\begin{align}
&\lambda(\hat{\mu}=1)=\lambda(2\pi T)\rightarrow \infty,
\label{lamda-infity-1}
\\
&\lambda(\hat{\mu}=2)\rightarrow \frac{4\pi^2}{\log 2},
\label{lamda-infity-2}
\end{align}
For $\hat{\mu}=1$, dividing both sides of Eqs.~(\ref{eq:gap-mq})--(\ref{eq:gap-ms}) by $\lambda$, and setting $\lambda\to\infty$ from Eq.~(\ref{lamda-infity-1}), we obtain
\begin{align}
\hat m_s
&= \frac{6\hat m_d+6.689253-3\log 2}
{12(\gamma_E+\log 2)},\nonumber\\
\hat m_q^2
&= \frac{1}{4}\Big[
1-\hat m_s+\hat m_s^2
+\hat m_d^2(3\gamma_E-7\log 2)-2\hat m_s\hat m_d
\nonumber\\
&\times(\gamma_E-\log 2)
+\hat m_d(1.830218+2\log\hat m_d)
\Big],\nonumber\\
&\hat m_d+\hat m_d^2(\gamma_E-\log 2)
-2\hat m_s\hat m_d
=1-\frac{1}{3}\hat m_s \nonumber\\ &\qquad\times
\left(-16.296185+4\gamma_E-12\log\hat m_s\right)
\nonumber\\
&\qquad-\hat m_s^2(1+12\log 2)
+8\hat m_q^2(\gamma_E+\log 2),
\end{align}
and by solving these three equations, we obtain 
\beq
\hat{m}_d=0.615263,\quad \hat{m}_s=0.544555,\quad \hat{m}_q=0.210265,
\label{mass-infinity-u1}
\eeq
which are finite constants. Since $\hat{m}_{n,{\rm LO}}\propto \sqrt{\lambda}$, $m_n/m_{n,{\rm LO}}=\hat{m}_n/\hat{m}_{n,{\rm LO}}\propto 1/\sqrt{\lambda(2\pi T)}\rightarrow 0$. This derivation can also apply to the case when setting $m_q=0$, which yields 
\beq
\hat{m}_d^{m_q=0}=0.851515,\quad \hat{m}_s^{m_q=0}=0.0286817.
\label{mass-infinity-u1-mq0}
\eeq

For $\hat{\mu}=2$, putting Eq.~(\ref{lamda-infity-2}) into Eqs.~(\ref{eq:gap-mq})--(\ref{eq:gap-ms}), and solving numerically, we obtain 
\beq
\hat{m}_d= 0.736868, \hat{m}_s =0.442212, \hat{m}_q = 0.253784,
\eeq
and by using Eqs.~(\ref{lamto0}) and (\ref{lamto01}), we obtain 
\beq
\frac{m_d}{m_{d,{\rm LO}}}=0.613483, \frac{m_s}{m_{s,{\rm LO}}}=0.520665, \frac{m_q}{m_{q,{\rm LO}}}=0.422578. 
\eeq

\bibliography{n2_HTLpt}

\end{document}